\documentclass[twocolumn]{aastex63}
\usepackage{rotating}
\usepackage{color, colortbl}
\usepackage{tikz}
\usepackage{afterpage}
\usepackage{pifont}
\usepackage[para]{threeparttable}
\usepackage{tabularx}
\usepackage{graphicx}

\shorttitle{Infant SNe II ZTF}
\shortauthors{Bruch et al.}

\graphicspath{{./}{figures/}}

\begin{document}

\title{The prevalence and influence of circumstellar material around hydrogen-rich supernova progenitors}

\correspondingauthor{Rachel J. Bruch}
\email{rachel.bruch@weizmann.ac.il}

\author[0000-0002-0786-7307]{Rachel J. Bruch}
\affil{Department of Particle Physics and Astrophysics 
Weizmann Institute of Science 
234 Herzl St.
76100 Rehovot, Israel}

\author{Avishay Gal-Yam}
\affil{Department of Particle Physics and Astrophysics 
Weizmann Institute of Science 
234 Herzl St.
76100 Rehovot, Israel}

\author{Ofer Yaron}
\affil{Department of Particle Physics and Astrophysics 
Weizmann Institute of Science 
234 Herzl St.
76100 Rehovot, Israel}

\author{Ping Chen}
\affil{Department of Particle Physics and Astrophysics 
Weizmann Institute of Science 
234 Herzl St.
76100 Rehovot, Israel}

\author[0000-0002-4667-6730]{Nora L. Strotjohann}
\affil{Department of Particle Physics and Astrophysics 
Weizmann Institute of Science 
234 Herzl St.
76100 Rehovot, Israel}

\author{Ido Irani}
\affil{Department of Particle Physics and Astrophysics 
Weizmann Institute of Science 
234 Herzl St.
76100 Rehovot, Israel}

\author{Erez Zimmerman}
\affil{Department of Particle Physics and Astrophysics 
Weizmann Institute of Science 
234 Herzl St.
76100 Rehovot, Israel}

\author[0000-0001-6797-1889]{Steve Schulze}
\affil{Department of Particle Physics and Astrophysics 
Weizmann Institute of Science 
234 Herzl St.
76100 Rehovot, Israel}
\affil{The Oskar Klein Centre, Department of Astronomy, Stockholm University, AlbaNova, SE-106 91 Stockholm, Sweden}

\author{Yi Yang}
\affil{Department of Particle Physics and Astrophysics 
Weizmann Institute of Science 
234 Herzl St.
76100 Rehovot, Israel}

\author[0000-0002-1031-0796]{Young-Lo Kim}
\affil{Universit\'e de Lyon, Universit\'e Claude Bernard Lyon 1, CNRS/IN2P3, IP2I Lyon, F-69622, Villeurbanne, France}
\affil{Department of Physics, Lancaster University, Lancs LA1 4YB, UK}

\author[0000-0002-8255-5127]{Mattia Bulla}
\affil{The Oskar Klein Centre, Department of Astronomy, Stockholm University, AlbaNova, SE-106 91 Stockholm, Sweden}

\author[0000-0003-1546-6615]{Jesper Sollerman}
\affil{The Oskar Klein Centre, Department of Astronomy, Stockholm University, AlbaNova, SE-106 91 Stockholm, Sweden}

\author{Mickael Rigault}
\affil{Universit\'e de Lyon, Universit\'e Claude Bernard Lyon 1, CNRS/IN2P3, IP2I Lyon, F-69622, Villeurbanne, France}

\author{Eran Ofek}
\affil{Department of Particle Physics and Astrophysics 
Weizmann Institute of Science 
234 Herzl St.
76100 Rehovot, Israel}

\author[0000-0001-6753-1488]{Maayane Soumagnac}
\affil{Department of Particle Physics and Astrophysics 
Weizmann Institute of Science 
234 Herzl St.
76100 Rehovot, Israel}
\affil{Physics Department, Bar Ilan University, Ramat Gan, Israel}

\author[0000-0002-8532-9395]{Frank J. Masci}
\affiliation{IPAC, California Institute of Technology, 1200 E. California Blvd, Pasadena, CA 91125, USA}


\author[0000-0002-4223-103X]{Christoffer Fremling}
\affil{Cahill Center for Astrophysics, California Institute of Technology, MC 249-17, 1200 E California Boulevard, Pasadena, CA, 91125, USA}

\author[0000-0001-8472-1996]{Daniel Perley}
\affil{Astrophysics Research Institute, Liverpool John Moores University, Liverpool Science Park, 146 Brownlow Hill, Liverpool L3 5RF, UK}

\author{Jakob Nordin}
\affil{Institute of Physics, Humboldt-Universit¨at zu Berlin, Newtonstr. 15, 12489 Berlin, Germany}

\author{S. Bradley Cenko}
\affil{Astrophysics Science Division, NASA Goddard Space Flight
Center, MC 661, Greenbelt, MD 20771, USA}
\affil{Joint Space-Science Institute, University of Maryland, College Park, MD 20742, USA}

\author[0000-0002-9017-3567]{Anna Y. Q.~Ho}
\affiliation{Department of Astronomy, Cornell University, Ithaca, NY 14853, USA}


\author{S. Adams}
\affil{Cahill Center for Astrophysics, California Institute of Technology, MC 249-17, 1200 E California Boulevard, Pasadena, CA, 91125, USA}

\author{Igor Adreoni}
\affil{Cahill Center for Astrophysics, California Institute of Technology, MC 249-17, 1200 E California Boulevard, Pasadena, CA, 91125, USA}

\author[0000-0001-8018-5348]{Eric C. Bellm}
\affil{DIRAC Institute, Department of Astronomy, University of Washington, 3910 15th Avenue NE, Seattle, WA 98195, USA}

\author[0000-0003-0901-1606]{Nadia Blagorodnova}
\affil{Department of Astrophysics/IMAPP, Radboud University, Nijmegen, The Netherlands}

\author{Kevin Burdge}
\affil{Cahill Center for Astrophysics, California Institute of Technology, MC 249-17, 1200 E California Boulevard, Pasadena, CA, 91125, USA}

\author[0000-0002-8989-0542]{Kishalay De}
\affil{Cahill Center for Astrophysics, California Institute of Technology, MC 249-17, 1200 E California Boulevard, Pasadena, CA, 91125, USA}

\author{Richard G. Dekany}
\affil{Caltech Optical Observatories, California Institute of Technology, MC 249-17, 1200 E California Boulevard, Pasadena, CA, 91125}

\author{Suhail Dhawan}
\affil{The Oskar Klein Centre, Department of Astronomy, Stockholm University, AlbaNova, SE-106 91 Stockholm, Sweden}

\author{Andrew J. Drake}
\affil{Division of Physics, Mathematics and Astronomy, California Institute of Technology, Pasadena, CA 91125, USA}

\author[0000-0001-5060-8733]{Dmitry A. Duev}
\affil{Division of Physics, Mathematics and Astronomy, California Institute of Technology, Pasadena, CA 91125, USA}

\author[0000-0002-3168-0139]{Matthew Graham}
\affil{Cahill Center for Astrophysics, California Institute of Technology, MC 249-17, 1200 E California Boulevard, Pasadena, CA, 91125, USA}

\author{Melissa L. Graham}
\affil{University of Washington, Department of Astronomy Box 351580 Seattle WA 98195-1580, USA}

\author{Jacob Jencson}
\affil{Cahill Center for Astrophysics, California Institute of Technology, MC 249-17, 1200 E California Boulevard, Pasadena, CA, 91125, USA}

\author{Emir Karamehmetoglu}
\affil{The Oskar Klein Centre, Department of Astronomy, Stockholm University, AlbaNova, SE-106 91 Stockholm, Sweden}
\affil{Department of Physics and Astronomy, Aarhus University, Ny Munkegade 120, DK-8000 Aarhus C, Denmark}

\author[0000-0002-5619-4938]{Mansi M. Kasliwal}
\affil{Cahill Center for Astrophysics, California Institute of Technology, MC 249-17, 1200 E California Boulevard, Pasadena, CA, 91125, USA}

\author[0000-0001-5390-8563]{Shrinivas Kulkarni}
\affil{Cahill Center for Astrophysics, California Institute of Technology, MC 249-17, 1200 E California Boulevard, Pasadena, CA, 91125, USA}

\author[0000-0001-9515-478X]{A.~A.~Miller}
\affiliation{Center for Interdisciplinary Exploration and Research in Astrophysics and Department of Physics and Astronomy, Northwestern University, 1800 Sherman Ave, Evanston, IL 60201, USA}
\affiliation{The Adler Planetarium, Chicago, IL 60605, USA}

\author[0000-0002-0466-1119]{James D. Neill}
\affil{Cahill Center for Astrophysics, California Institute of Technology, MC 249-17, 1200 E California Boulevard, Pasadena, CA, 91125, USA}

\author{Thomas A. Prince}
\affil{Division of Physics, Mathematics and Astronomy, California Institute of Technology, Pasadena, CA 91125, USA}

\author{Reed Riddle}
\affil{Caltech Optical Observatories, California Institute of Technology, MC 249-17, 1200 E California Boulevard, Pasadena, CA, 91125}

\author{Benjamin Rusholme}
\affil{IPAC, California Institute of Technology, 1200 E. California Blvd, Pasadena, CA 91125, USA}

\author{Y. Sharma}
\affil{Cahill Center for Astrophysics, California Institute of Technology, MC 249-17, 1200 E California Boulevard, Pasadena, CA, 91125, USA}

\author{Roger Smith}
\affil{Caltech Optical Observatories, California Institute of Technology, MC 249-17, 1200 E California Boulevard, Pasadena, CA, 91125}

\author{Niharika Sravan}
\affil{Division of Physics, Mathematics, and Astronomy, California Institute of Technology, Pasadena, CA 91125, USA}

\author[0000-0002-5748-4558]{Kirsty Taggart}
\affil{Astrophysics Research Institute, Liverpool John Moores University, Liverpool Science Park, 146 Brownlow Hill, Liverpool L3 5RF, UK}

\author{Richard Walters}
\affil{Caltech Optical Observatories, California Institute of Technology, MC 249-17, 1200 E California Boulevard, Pasadena, CA, 91125}

\author{Lin Yan}
\affil{Cahill Center for Astrophysics, California Institute of Technology, MC 249-17, 1200 E California Boulevard, Pasadena, CA, 91125, USA}

\begin{abstract}

Narrow transient emission lines (flash-ionization features) in early supernova (SN) spectra trace the presence of circumstellar material (CSM) around the massive progenitor stars of core-collapse SNe. The lines disappear within days after the SN explosion, suggesting that this material is spatially confined, and originates from enhanced mass loss shortly (months to a few years) prior to explosion. We performed a systematic survey of H-rich (Type II) SNe discovered within less than two days from explosion during the first phase of the Zwicky Transient Facility (ZTF) survey (2018-2020), finding thirty events for which a first spectrum was obtained within $< 2$ days from explosion. The measured fraction of events showing flash ionisation features ($>36\%$ at $95\%$ confidence level) confirms that elevated mass loss in massive stars prior to SN explosion is common. We find that SNe II showing flash ionisation features are not significantly brighter, nor bluer, nor more slowly rising than those without. This implies that CSM interaction does not contribute significantly to their early continuum emission, and that the CSM is likely optically thin. We measured the persistence duration of flash ionisation emission and find that most SNe show flash features for $\approx 5 $ days. Rarer events, with persistence timescales $>10$ days, are brighter and rise longer, suggesting these may be intermediate between regular SNe II and strongly-interacting SNe IIn.

\end{abstract}



\section{Introduction}
Early observations of Type II supernovae (SNe II) reveal that a large fraction shows transient narrow emission lines of highly-ionised species (\citealt{khazov2016, bruch2021}). Such lines may result either from the recombination of slowly expanding circumstellar medium (CSM) excited and ionised by the SN shock breakout and the shock-cooling emission \citep{yaron2017, galyam2014, niemela1985}; or from the recombination of unshocked CSM excited by radiation originating from shocks driven by underlying ejecta-CSM interaction. The former excitation mechanism is called flash ionisation \citep{galyam2014}, while the latter would better be described by shock ionisation \citep{terreran2022}. 
After a few days, in both cases, these lines disappear, which suggests that the CSM is confined to a small volume around the progenitor, and is swept up by the ejecta.

Follow-up observations (e.g. rapid-response spectroscopy and multiband photometry, \citealt{galyam2011}) are useful to probe the properties of the progenitor and its surroundings. For example, \citet{rabinak2011} (RW11, hereafter) motivate acquiring daily multiband photometry at early times to constrain the radius of the progenitor as well as the explosion energy per unit mass. The RW11 model is no longer valid once the recombination phase has started, which is marked by the emergence of broad hydrogen P-Cygni-like lines. Rapid-response spectroscopy with a day cadence is hence needed to identify the beginning of the recombination phase. As we show below, detecting narrow emission lines at early time is indicative of interaction with CSM, which could also invalidate the use of such models. Studies of large samples of such events are important to establish the characteristic SN progenitor channels and the conditions which bring them to explosion. 

Signatures of a dense and extended CSM are observed in Type IIn SNe. These are hydrogen rich SNe that show strong and narrow Balmer emission lines for an extended period of time and do not develop broad hydrogen features, characteristic of the high expansion velocity of the SN ejecta around peak light (\citealt{schlegel1990,filippenko1997,kiewe2012, Smith2014}). The narrow features come from slowly expanding CSM, energized from within by the shock interaction between the expanding ejecta and dense CSM, (\citealt{chugai1994}). Such interaction can last from weeks to years after the explosion. 
Photometrically, SNe IIn are typically brighter (M$_{peak,\text{r}}\approx -19 $ mag) and can rise to peak on longer timescales ($t_{rise,\text{r}} > 20 $d) than SNe II (M$_{peak, \text{r}} \approx -17.5$ mag and $t_{rise,\text{r}} < 20 $d, respectively), see \cite{nyholm2020} for SNe IIn and \cite{rubin2016_Wax} for SNe II. This extra luminosity is thought to come from the interaction with the CSM. Indeed, the kinetic energy from the ejecta is converted to X-Rays via collisionless shocks in the CSM, and then converted to visible light if enough optical depth is present. 

The origin of this CSM is usually attributed to an elevated mass loss prior to the explosion. Indeed, massive stars are known to experience mass loss throughout their lives ($\dot{M}< 10^{-4} M_{\odot}$.year$^{-1}$ for red supergiants (RSG), see chapter 8 in \cite{prialnik2009} and Figure 3 in \cite{Smith2016}). However, there is also evidence for episodic elevated mass-loss just prior to the explosion  \citep{ofek2013,ofek2014,Strotjohann2021,jacobson2022}. Such precursor emission with enhanced mass-loss could originate from strong convection close to the core, in the late stages of nuclear burning, which generates waves that heat the stellar envelope intensively \citep{quataert2012,Shiode2014}. Other ideas involve for example sudden energy release in deep layers of the stars from late-stage nuclear burning instabilities \citep{meakin2007}. For SNe IIn, the CSM distribution requires extensive mass loss over a long period of time (years prior explosion, see \cite{Strotjohann2021,ofek2013,ofek2014}). In the case of flash features, the CSM presumably lies in a more confined space and could result from mass-loss episodes, occurring shortly\footnote{Months to weeks.} prior to the explosion \citep{galyam2014,yaron2017,jacobson2022}.

Since CSM shock interaction seems to power the light curve of SNe IIn, it is possible that CSM interaction contributes also to the early light curve of SNe II with a confined CSM shell, \citep{morozova2017}. Thus we expect that SNe II with CSM reach higher luminosities than those without CSM, as hypothesised in \cite{Hosseinzadeh2018}.\\

We present here a systematic search for flash-ionisation features in hydrogen-rich SNe shortly after explosion ($<2.5$\, days). In Section \ref{obs}, we describe our construction of a large sample of infant SNe and our observations. In section \ref{analyse}, we present our analysis, and thus discuss our results in section \ref{resultsdiscuss}. We conclude in section \ref{conclusion}.

In this paper, we assume cosmology parameters from \citealt{Planck2013}, which yields a Hubble constant at  $H_0 = 67.3 \pm 1.2\,$ km.Mpc$^{-1}$.s$^{-1}$. Magnitudes are given in the AB system. The photometry and spectra presented in this paper will be made public through WiseRep \footnote{https://www.wiserep.org/} (\citealt{wiserep2012}).

\begin{figure*}
    \hspace{-2cm}
    \includegraphics[width=1.2\textwidth]{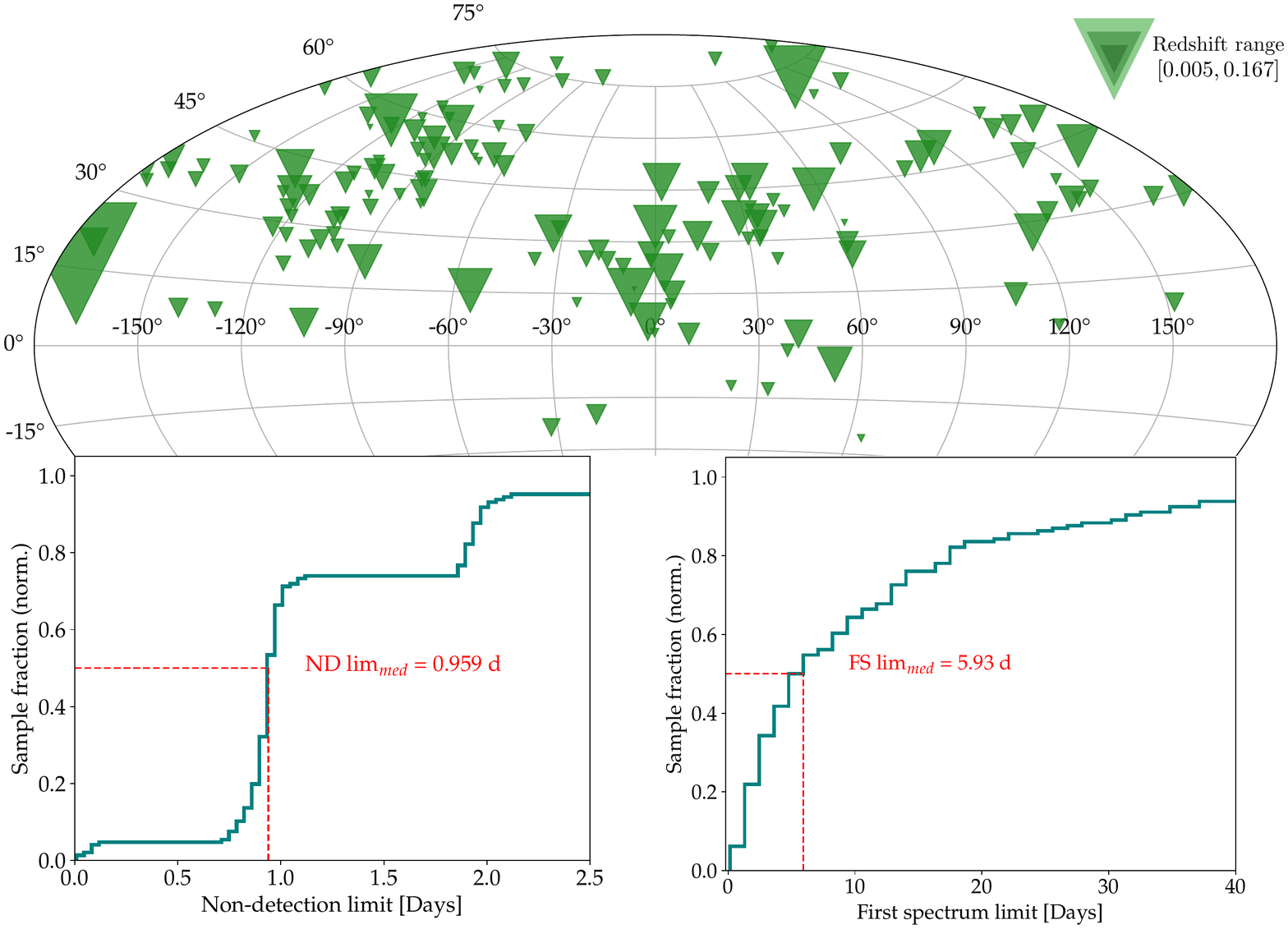}
    \caption{Sample statistics: celestial distribution of the 148 candidates of the real-infant sample. Large triangles designate low redshift events, smaller ones higher redshift. Our sample ranges from $z=0.005$ to $z=0.167$        \textit{Bottom left:} Non-detection limit distribution. Half of the sample was discovered within $0.9\,\text{days}$ of the last non-detection. \textit{Bottom right:} Time of the first spectrum. 22\% of the sample has a first spectrum within less than 2.5d the last non-detection.}
    \label{fig:sample_stats}
\end{figure*}

\section{Observations} \label{obs}

\subsection{Sample construction}
We constructed our sample through both real-time data scanning and complementary archival search. 
We consider SNe detected and classified by ZTF-I, i.e. between March 2018 and December 2020, and restrict our search to spectroscopically-classified SN II, IIn and IIb. A candidate is considered spectroscopically classified if at least one spectrum exists on the ZTF Growth Marshal \citep{Kasliwal2019}. To constrain non-detection limits, we consider exclusively the light curves provided by P48. 
We compute the non-detection limits from the light curves of 1252 spectroscopically-classified hydrogen rich SNe\footnote{SNe II, IIb and IIn} and find that 425 have a non-detection within $<2.5$ days from the first detection. We carried out forced photometry for the candidate sample composed of these 425 H-rich candidates. At this stage, the light curves were not corrected for extinction or redshift. \\
A candidate is qualified as a "Real Infant" SN if for a given filter (r and/or g) there is a non-detection within $<2.5$ days, and if the magnitude rise above the limiting magnitude of this non-detection is $>0.5$ magnitudes, as described in \cite{bruch2021}. \\
Forced photometry is more sensitive than the ZTF alert detection pipeline (see \citealt{yao2019}) and some initially reported non-detections turn into faint detections (e.g.: SN 2019eoh) and reveal an intra-night rise $> 0.5$ mag. Such fast rising behaviour is characteristic of an infant SN and we therefore also add events with an intra-night rise of at least 0.5 magnitudes in the same band to our sample. 

Some SNe were mistakenly classified as Type IIn SNe, based on early spectra where narrow hydrogen lines could be flash ionisation lines. Some candidates, such as SN 2019qch, were classified as a SN IIn based on long lasting flash ionisation features. This SN developed a classic broad P-Cygni profile, characteristic of spectroscopically normal SNe II later. Whenever an event was classified as a SNe IIn, but showed this latter  behaviour, we changed the classification to SN II.\\

The final sample (see Tables \ref{tab:allSNII},\ref{tab:allSNIIn},\ref{tab:allSNIIb} in the Annex) is composed of 148 SNe discovered between March 2018 and December 2020.  The median non-detection limit from the first detection is 0.9 days. The median acquisition of a first spectrum from the last non detection is $\approx 6$ days, see Figure \ref{fig:sample_stats}.

\subsection{Photometry} 
\paragraph{Alert system photometry from P48} 
We had access to the partnership and public photometric data stream from ZTF \citep{bellmztf2019,grahamztf2019,dekany2020}. We used this pipeline and its real-time alert distribution via the GROWTH marshal to look for infant candidates on a daily basis. We complementary used the AMPEL system for filtering the incoming alerts, \citep{nordin2019,soumagnac2018}. 

\paragraph{Real time forced photometry bot} 
During the scanning campaign, we also used the \texttt{fpbot} \citep{reusch2022}. The fpbot returns in real time the forced photometry of a ZTF object and sends it to the platform \texttt{Slack}. Such a service was useful upon the discovery of a candidate to ensure that the first detection was real prior to triggering spectroscopic follow up. 

\paragraph{Forced photometry} 
We carried out forced photometry for the entirety of our sample and applied the quality cuts described in \cite{bruch2021}. 
We  visually inspect the resulting magnitude light curves from forced photometry as well as the alert system photometry to determine the first detection and last non-detection. We retrieved pre-discovery detections from the forced photometry magnitude light curves for 53 candidates \footnote{SNe 2018cfj, 2018ccp, 2018dfa, 2018cyh, 2018cxn, 2018cug, 2018lti, 2018fzn, 2018fif, 2018fpb, 2018fso, 2018iwe, 2018gfx, 2018gts, 2018iug, 2018iua, 2019cem, 2019eoh, 2019ewb, 2019fkl, 2019hln, 2019mge, 2019lkw, 2019aaqx, 2019pdm, 2019njv, 2019oba, 2019oot, 2019pgu, 2019ozf, 2019qch, 2019rsw, 2019smj, 2019tbq, 2019vdl, 2020ks, 2020cnv, 2020iez, 2020lcc, 2020oco, 2020ovk, 2020pnn, 2020pvg, 2020rsc, 2020smm, 2020ufx, 2020uim, 2020uhf, 2020uqx, 2020urc, 2020xkx, 2020ykb, 2020yyo}. The median value is 0.99 days prior to the detection from the alert system. From this point on,  we use the first detection and last non-detection from the forced-photometry light curves. 
Some candidates from 2020 have measurements from Caltech-partnership data stream. Since we did not have access to them at the time of the analysis, they are not included in this study. This does not impact significantly our analysis. It was established in \cite{bellmztf2019} that the ZTF single image limit is on average 21 magnitude. We hence remove any detections from the forced photometry which are below 21st magnitude. The forced-photometry light curves can be found in Table \ref{tab:fullforcephot}.

\begin{table*}
\caption{Forced-photometry light curves of the 148 events in our Real Infant sample. The full version can be found online. }
\hspace{-1.5cm}
\begin{tabular}{cccccccccc}
\hline
Time from & Time & Flux & Flux & Apparent & $\delta$m  & Absolute  & $\delta$M & Filter & ZTF  \\
EED      &       &      & error& magnitude&            & magnitude &           &        &  name \\
         &   [JD]& $10^{-8}$[Mgy]&$10^{-8}$[Mgy]&    (m)    &           &    (M)      &           &        &  \\
\hline
\hline
... & ... & ... & ... & ... & ... & ... & ... & ... & ... \\
17.420 & 2458898.817 & 4.97484548 & 0.16902504 & 18.26 & 0.04 & -17.74 & 0.04 & r & ZTF18aaaibml \\
18.404 & 2458899.801 & 5.09761181 & 0.14810007 & 18.23 & 0.03 & -17.76 & 0.03 & r & ZTF18aaaibml \\
20.478 & 2458901.875 & 4.67906920 & 0.13268852 & 18.32 & 0.03 & -17.67 & 0.03 & r & ZTF18aaaibml \\
22.413 & 2458903.810 & 4.40633990 & 0.12260471 & 18.39 & 0.03 & -17.60 & 0.03 & g & ZTF18aaaibml \\
23.330 & 2458904.727 & 5.04859332 & 0.15628758 & 18.24 & 0.03 & -17.75 & 0.03 & r & ZTF18aaaibml \\
23.377 & 2458904.774 & 3.85174173 & 0.10942362 & 18.54 & 0.03 & -17.46 & 0.03 & g & ZTF18aaaibml \\
... & ... & ... & ... & ... & ... & ... & ... & ... & ... \\
\hline
\end{tabular}
\tablecomments{$\delta$m and $\delta$M are respectively the error on the apparent and absolute magnitude. This table includes the flux measurements returned by the forced photometry pipeline, and the time from the estimated explosion date (EED).}
\label{tab:fullforcephot}

\end{table*}

\subsection{Spectroscopy}

Our goal was to obtain rapid spectroscopy of infant SN candidates following the methods of \cite{galyam2011}. This was made possible using 
rapid ToO follow-up programs as well as on-request access to scheduled nights on various telescopes. During the active search for new transients, we applied the following criteria for rapid spectroscopic triggers: The robotic SEDm (see below) was triggered for all candidates brighter than a magnitude threshold of $19$\,mag. The co-location of the P60 and ZTF/P48 on the same mountain, as well as the P60 robotic response capability, enable rapid, often same-night, response to ZTF events. However, the low resolution ($R\sim100$) of the instrument limits our capability to characterise narrow emission lines. This, along with the overall sensitivity of the system, motivated us to obtain higher-resolution follow-up spectroscopy with larger telescopes, particularly for all infant SNe fainter than $r\sim19$ mag at discovery.  Higher-resolution spectra (using WHT, Gemini, or other available instruments) were triggered for events assured to be of extragalactic nature \footnote{We crosschecked the position of the alerts with known catalogues such as VIZIER, \cite{Ochsenbein2000}}, showing recent non-detection limits (within $2.5$\,d prior to first detection) as well as a significant brightening compared to a recent non-detection. \\

We present here the spectroscopic facilities we used during our search for infant SNe II. 

\paragraph{P60/SEDm} The Spectral Energy Distribution Machine (SEDm; \cite{benami2012,blagorodnova2018,neillSEDm2019}) is a high-throughput, low-resolution spectrograph mounted on the 60" robotic telescope (P60; \cite{cenko2006}) at Palomar observatory. $65\%$ of the time on the SEDm was  dedicated to ZTF partnership follow up. SEDm data are reduced using an automated pipeline \citep{rigaultsedm2019,Kim2022}. \\

\paragraph{LT/SPRAT} We used the Spectrograph for the Rapid Acquisition of Transients (SPRAT; \cite{Piascik2014}). It is a highthroughput, low-resolution spectrograph mounted on the Liverpool Telescope (LT; 58), a 2 meter robotic telescope at the Observatorio del Roque de Los Muchachos in Spain. All the spectra were reduced using the standard pipeline provided by the observatory.\\

\paragraph{P200/DBSP} We used the Double Beam SPectrograph (DBSP; \cite{DBSP1982}) mounted on the 5m Hale telescope at Palomar Observatory (P200) to obtain follow-up spectroscopy in either ToO mode or during  classically-scheduled nights. The default configuration used the 600/4000 grism on the blue side, the 316/7150 grating on the red side, along with the D55 dichroic, achieving a spectral resolution $R\sim 1000$. Spectra obtained with DBSP were reduced using the pyraf-dbsp pipeline \citep{bellm2016}. \\

\paragraph{WHT/ISIS\&ACAM} We obtained access to the 4.2m William Herschel Telescope (WHT) at the Observatorio del Roque de los Muchachos in La Palma, Spain, via the Optical Infrared Coordination Network for Astronomy (OPTICON\footnote{https://www.astro-opticon.org/index.html}) program\footnote{Program IDs OPT/2017B/053, OPT/2018B/011, OPT/2019A/024, PI Gal-Yam}. We used both single-slit spectrographs ISIS and ACAM \citep{ACAM2008} in ToO service observing mode. The delivered resolutions were $R\sim1000$ and $R\sim400$, respectively. Spectral data were reduced using standard routines within IRAF \footnote{{IRAF} \textbf{was} distributed by the National Optical Astronomy Observatories, which are operated by the Association of Universities for Research in Astronomy, Inc., under cooperative agreement with the National Science Foundation. }.\\

\paragraph{Keck/LRIS} We used the Low-Resolution Imaging Spectrometer (LRIS; \cite{Keck1995}) mounted on the Keck-I 10m telescope at the W. M. Keck Observatory in Hawaii in either ToO mode or during scheduled nights. The data were reduced using the LRIS automated reduction pipeline Lpipe \citep{perley2019}.\\

\paragraph{GMOS/Gemini} We used the Gemini Multi-Object Spectrograph (GMOS; \citealt{Hook2004}) mounted on the Gemini North 8m telescope at the Gemini Observatory on Mauna Kea, Hawaii. All observations were conducted at a small airmass ($\lesssim1.2$). For each SN, we obtained 2$\times$900\,s exposures using the B600 grating with central wavelengths of $520$\,nm and $525$\,nm. The $5$\,nm shift in the effective central wavelength was applied to cover the chip gap, yielding a total integration time of $3600$\,s. A 1.0$\arcsec$-wide slit was placed on each target at the parallactic angle. The GMOS data were reduced following standard procedures using the Gemini IRAF package.  \\

\paragraph{ARC/DIS} We used the Dual Imaging Spectrograph (DIS) on the Astrophysical Research Consortium (ARC) 3.5\,m telescope at Apache Point Observatory (APO) during scheduled nights. The data were reduced using standard procedures and calibrated to a standard star obtained on the same night using the PyDIS package \citep{davenport}.\\

\paragraph{NOT/ALFOSC} Some data presented here were obtained with the Andalucia Faint Object Spectrograph and Camera (ALFOSC) mounted on the 2.56\,meter Nordic Optical Telescope (NOT)\\

\paragraph{VLT/FORS2} Some data presented here were obtained with the FOcal Reducer/low dispersion Spectrograph 2 on the Very Large Telescope in Long Slit spectroscopic mode. The data were obtained as part of the adH0cc\footnote{https://adh0cc.github.io/} project, based on the ESO-VLT Large Programme 1104.A-0380.\\

\paragraph{VLT/X-Shooter} Some data presented here were obtained with X-Shooter (\cite{Vernet2011}) as part of the ToO programme for Infant Supernovae. Optical spectra were obtained with the VIS spectrograph with an exposure time of $1450$s.\\ 
The data were reduced following \citet{Selsing2019a}. In brief, we first removed cosmic-rays with the tool  \texttt{astroscrappy}\footnote{\href{https://github.com/astropy/astroscrappy}{https://github.com/astropy/astroscrappy}}, which is based on the cosmic-ray removal algorithm by \citet{vanDokkum2001a}. Afterwards, the data were processed with the X-shooter pipeline v3.3.5 and the ESO workflow engine ESOReflex \citep{Goldoni2006a, Modigliani2010a}. The UVB and VIS-arm data were reduced in stare mode to boost the S/N by a factor of $\sqrt{2}$ compared to the standard nodding mode reduction. The individual rectified, wavelength- and flux-calibrated two-dimensional spectra files were co-added using tools developed by J. Selsing\footnote{\href{https://github.com/jselsing/XSGRB_reduction_scripts}{https://github.com/jselsing/XSGRB\_reduction\_scripts}}. 

The classification spectra of each SN can be found on the Transient Name Server\footnote{https://www.wis-tns.org/}. \\

\section{Analysis} \label{analyse}



Since our goal is to compare the photometric and spectroscopic behaviour of SNe II showing flash features and SNe II that do not, we need to estimate the explosion time, peak magnitude and rise time for each SN. We also discuss the identification of flash ionisation features in early spectra of SNe II.

\subsection{Explosion time estimation}

We use the fitter \texttt{iminuit} (\citealt{iminuit}) to fit the empirical function $f(t) = a \times (t-t_{exp})^n$, where $f(t)$ is the SN flux and $t_{exp}$ the time of zero flux. For each SN, we fit data from $-10$ until $2.5$, $3.5$, $4.5$ and $5.5$ days from first detection in $r$ and $g$ band. We inspect each fit visually. Whenever more than two fits (per band) were of poor quality, we adopted the time of zero flux as the mid point between the last non-detection limit and the first detection  \footnote{2020buc, 2018clq, 2018dfi, 2018iua, 2018iuq, 2018jak, 2019dvw, 2019ehk, 2019fkl, 2019fmv, 2019kes, 2019mge, 2019mor, 2019aaqx, 2019oot, 2019pkh, 2019rwd, 2019smj, 2019vdl, 2020cnv, 2020drl, 2020dya, 2020jmb, 2020jww, 2020pvg, 2020rhg, 2020qvw, 2020rfs, 2020rth, 2020sfy, 2020sbw, 2020sjv, 2020sur, 2020ult, 2020umi, 2020urc, 2020xkx}. We consider fits poor if they do not converge or if there are fewer than three observations after the first detection. 
We measure here the observer's time of zero flux and present it as the estimated explosion date (EED).

\subsection{Peak magnitude  estimation}
Using the methods described in \cite{bruch2021}, we estimate the peak magnitude: we correct the light curve for Milky Way extinction and redshift and fit a third degree polynomial to the light curve around the visual maximum. We repeat the fitting procedure 100 times and randomly vary the start and end dates of the fit. We estimate the value of the peak as the median of the maximum values. The rise time is the time from the estimated explosion date (EED) to the measured maximum. The error on the absolute peak magnitude is:

\begin{equation}
    \delta M_{peak} = \sqrt{\delta m_{peak}^2 + \delta \mu^2}
\end{equation}

with $\delta m_{peak}$, the standard deviation on the peak magnitude measurements and $\delta\mu$ is the error on the distance modulus, calculated from the error on the redshift as: $\delta\mu =\frac{ 5\delta z}{ln(10)\times z}$.


\subsection{Light curve interpolation}

\begin{figure}[h!]
\hspace{-2cm}

    \includegraphics[trim = 20 12 20 60,clip,width=1.1\columnwidth]{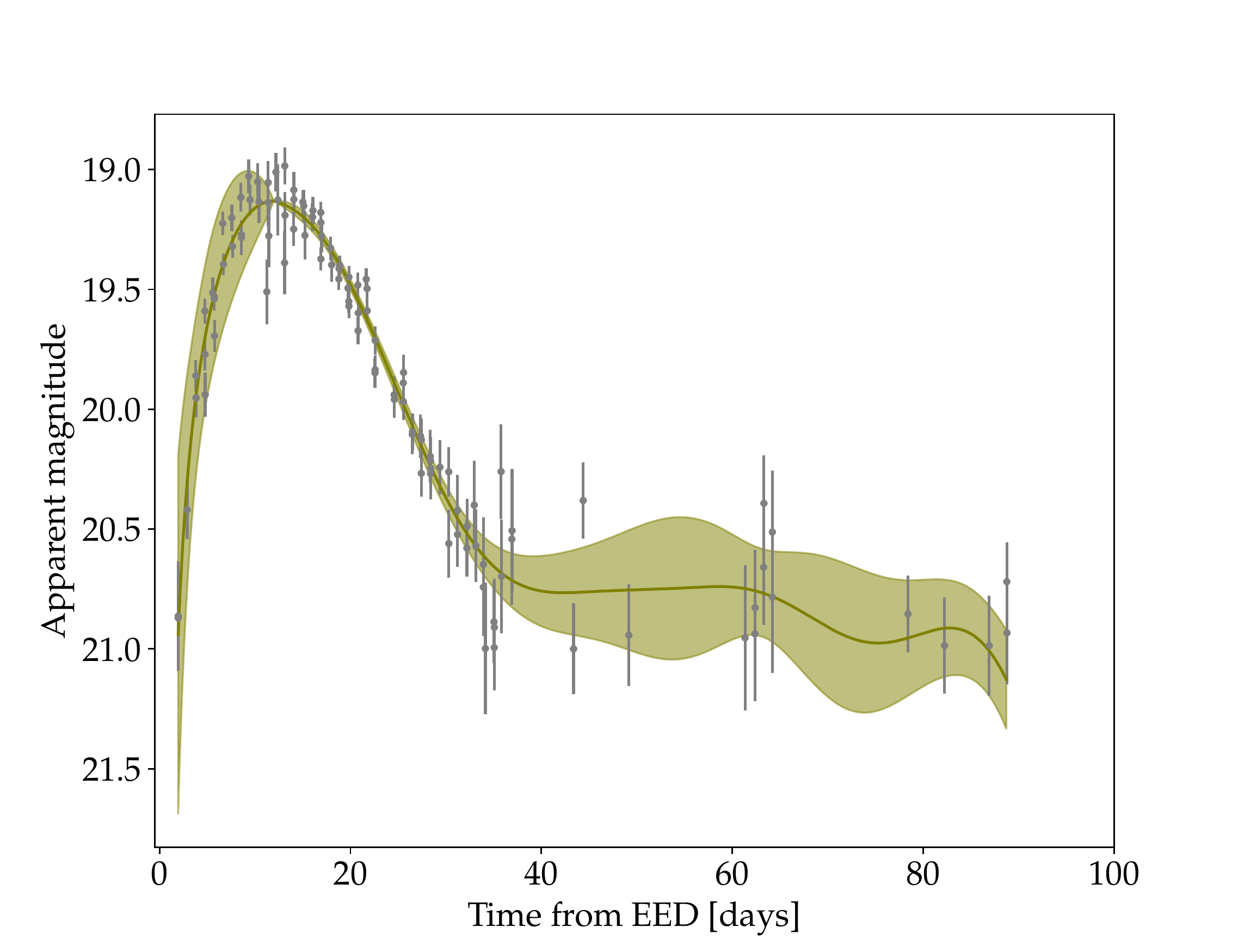}
    \hspace{-2cm}
    \includegraphics[trim = 20 12 20 60,clip,width=1.1\columnwidth]{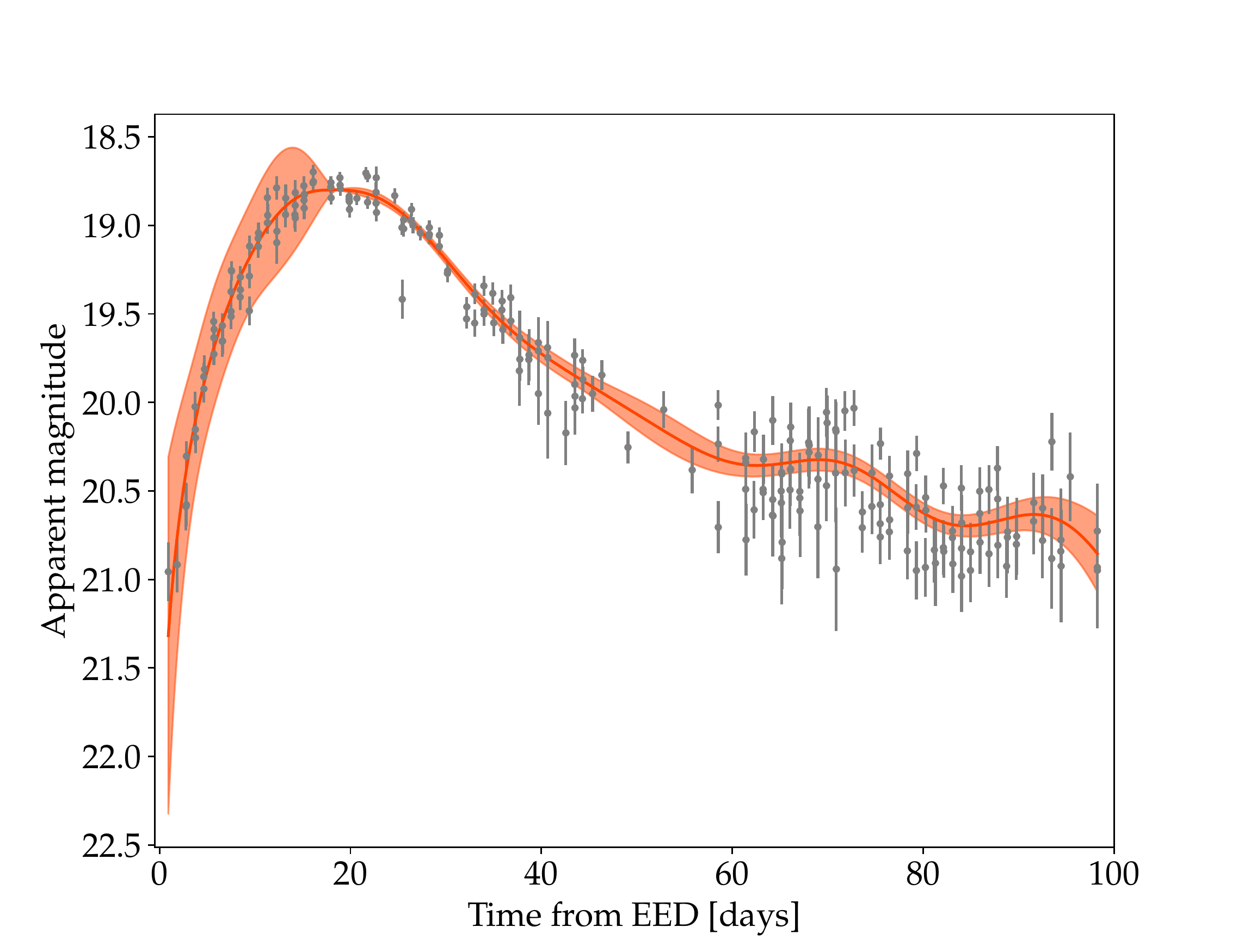}
    \caption{Full lightcurve interpolation of e.g. ZTF18abcezmh in the $g$ band (up) and $r$ band (down). The colored band represents the error on the interpolation from either the early time fit or the Gaussian process interpolation.}
    \label{fig:fullLCinterp}
\end{figure}

We also want to compare the colour at peak of each SN. We hence need to interpolate the light curve in each band. 
We divide the light curve into two parts: the rise and the decline. The rise is fit by either a simple exponential law or a broken power law until the estimated peak flux.  The decline is fit using gaussian processes. We use the forced-photometry light curves corrected for galactic extinction and redshift. \\

The rising lightcurve usually shows a fast rise followed by a slower rise. We use either a simple exponential law or a broken power law inspired by Equation (2) in \cite{johannesson2006}. Indeed, a simple exponential law is not enough in some cases to fit the full rise from EED to peak. The best fit is determined by a $\chi^2$ test. We fit the rise in flux space. Both functions are bounded to the estimated peak flux at the estimated peak time. The simple power law can be written as: 
\begin{equation}
     F = - A  \left(-\left(t-t_{peak}\right)\right)^n + F_{peak}
\end{equation}
 The broken power law is given by:
\begin{equation}
    F = -A\left( \left(\frac{-(t-t_{peak})}{t_{break}}\right)^{-\alpha_1} + \left(\frac{-(t-t_{peak})}{t_{break}}\right)^{-\alpha_2}\right) + F_{peak}
\end{equation}

Where $F_{peak}$, $t_{peak}$ are respectively the peak flux and rise time, and $t_{break}$ is the time where the transition between the two power laws happens. We use the \texttt{Minuit} optimiser, based on a least-square test, to choose which law fits best in each case. We then convert the obtained interpolation from flux space to magnitude space.\\

We use Gaussian processes to interpolate the decline part of the lightcurve. 
In order to estimate the overall decrease of magnitude per day after peak,  we fit a linear function to the light curve after the estimated peak : 
\begin{equation}
    mag = a \times (t - t_{peak}) + m_{peak}
\end{equation}
We exclude data points below 21 st magnitude and visually select the end of the region we choose to fit (usually $\approx 40$ days), usually corresponding to the last measurements before the SN is not observable anymore. 
We use this estimated linear decline curve as the mean function for Gaussian processes interpolation.

 Gaussian processes are not suitable to interpolate early SN light curves because they require a kernel that quantifies on what time scales the entire light curve varies. Indeed, Gaussian processes utilise a set of priors on the characteristic behaviour of the data. These priors are encapsulated in the Kernel from which each random function is drawn. One of the most basic assumptions for the Kernel is the characteristic size and amplitude of variation, i.e. two datapoints separated by length \textit{x} have a correlated behaviour and can vary over \textit{A} range of amplitude.  For us, the characteristic length for two points to behave alike is time, and the amplitude is some range of magnitude. From explosion to peak, very young Cor-Collapse SNe (CC SNe) first rise within hours, but, as the ejecta have expanded to a larger radius, they then vary over much longer time scales. There is therefore no single characteristic time scale for this phase, and hence for the full lightcurve.
However, in the linear decline phase from peak, we can assume that the characteristic timescale throughout the decline phase \footnote{Prior to the fall from the plateau and $^{56}$Co decay tail} is almost constant. We choose this time to be $\tau \approx 100$ days. The full light curve interpolation is stitched together at the estimated peak mag and time. See Figure \ref{fig:fullLCinterp} for an example.
 
 \begin{figure*}
    \centering
    \includegraphics[trim = 20 50 20 75,clip,width=\columnwidth]{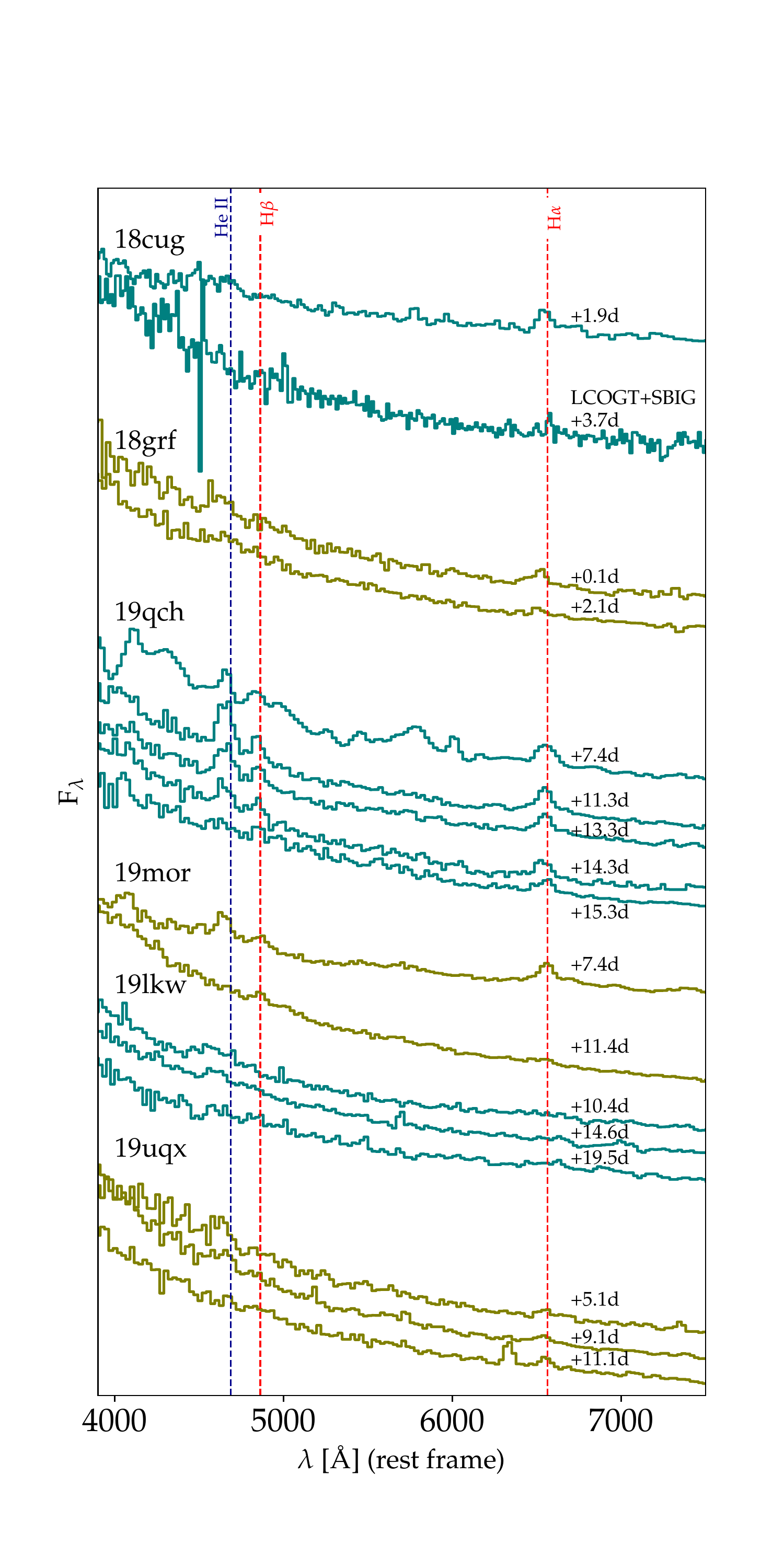}
    \includegraphics[trim = 20 50 20 75,clip,width=\columnwidth]{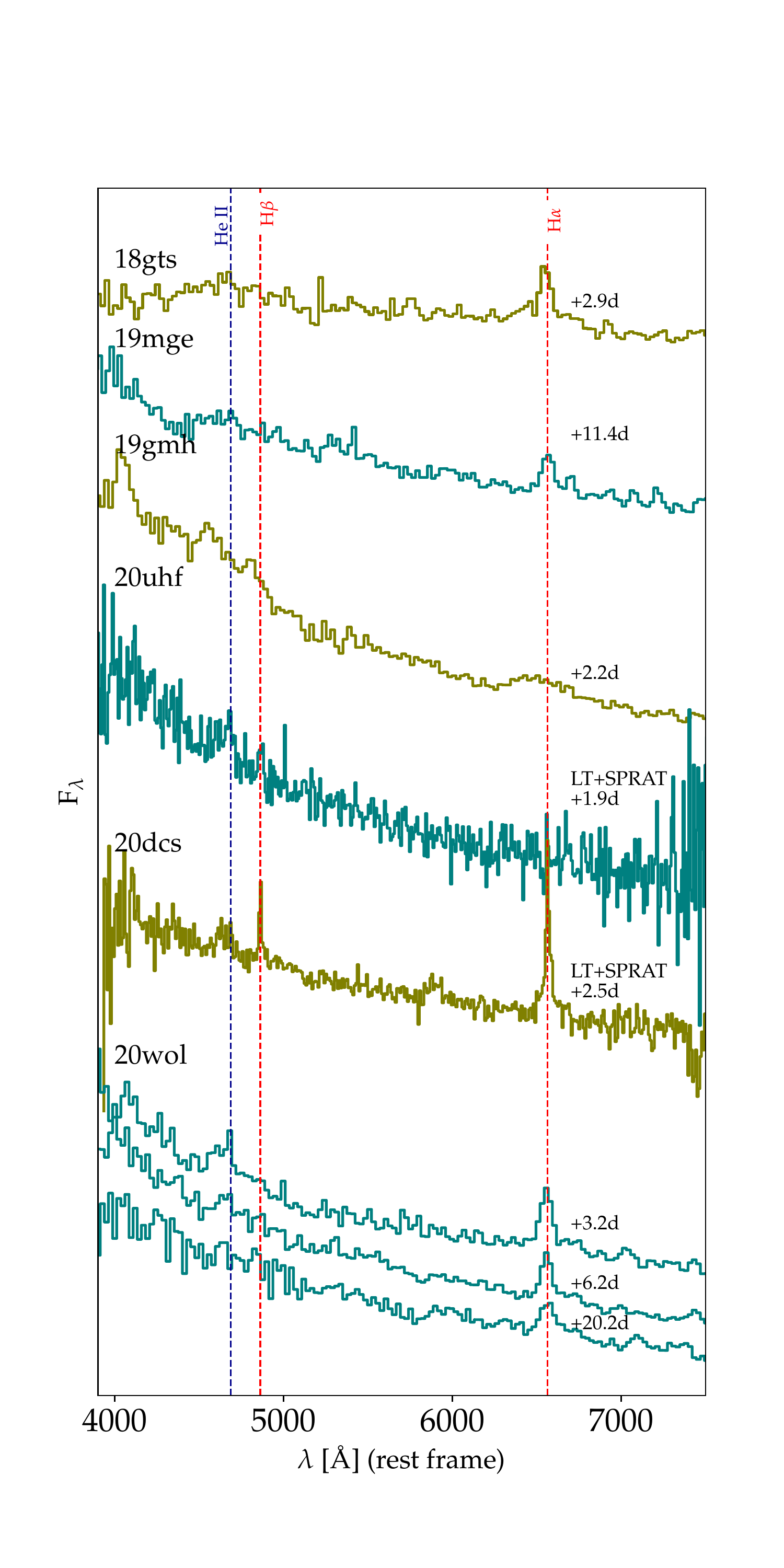}
    \caption{Collection of flash ionisation features spectra for the low-resolution group. Unless written otherwise, spectra were observed with SEDm+P60. The times indicated on the red side of the spectra correspond to the time of acquisition of the spectra from the estimated explosion date.}
    \label{fig:lowresflash}
\end{figure*}

\begin{figure*}[t!]
    \centering
    \includegraphics[trim = 20 50 20 75,clip,width=\columnwidth]{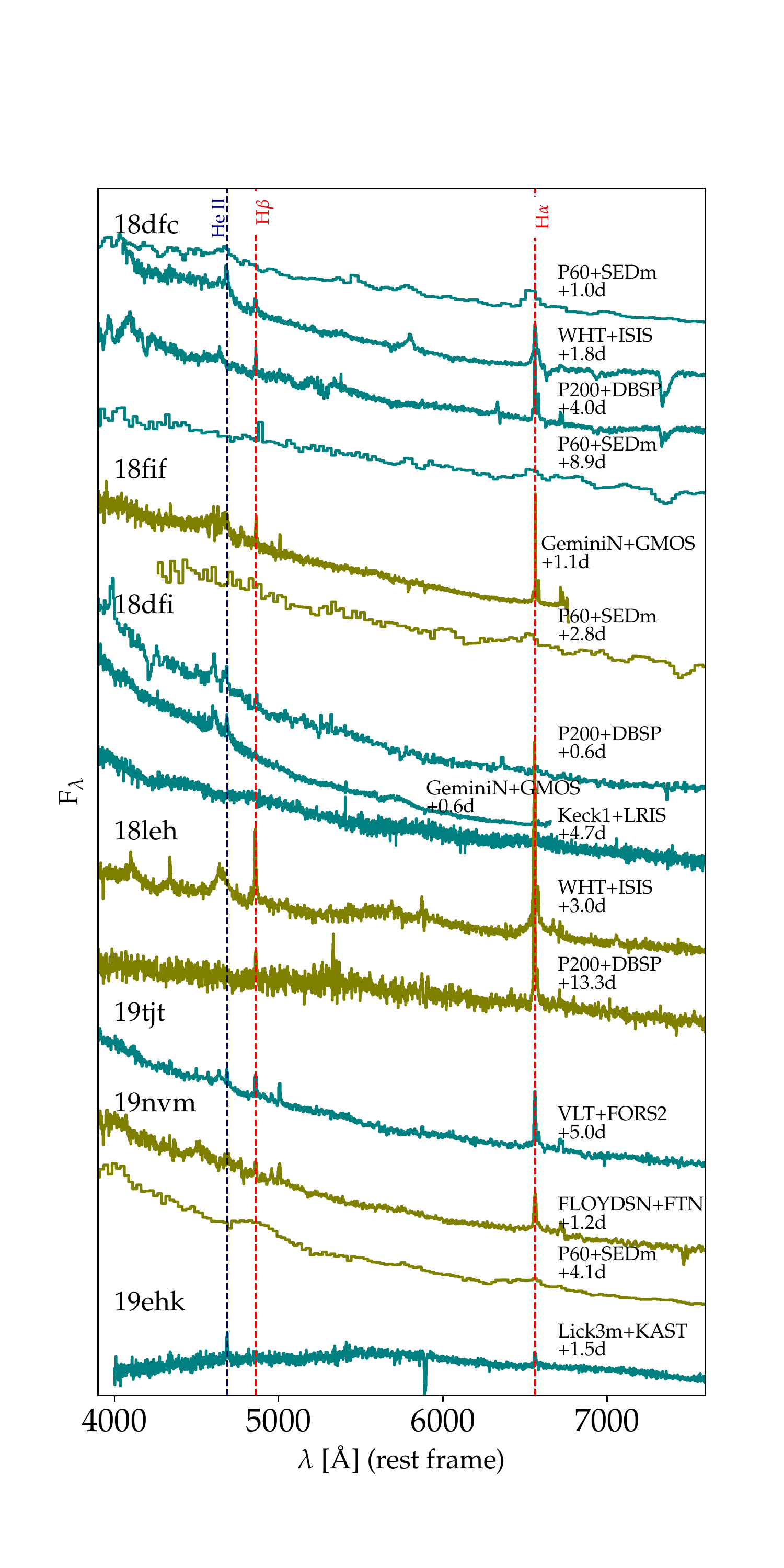}
    \includegraphics[trim = 20 50 20 75,clip,width=\columnwidth]{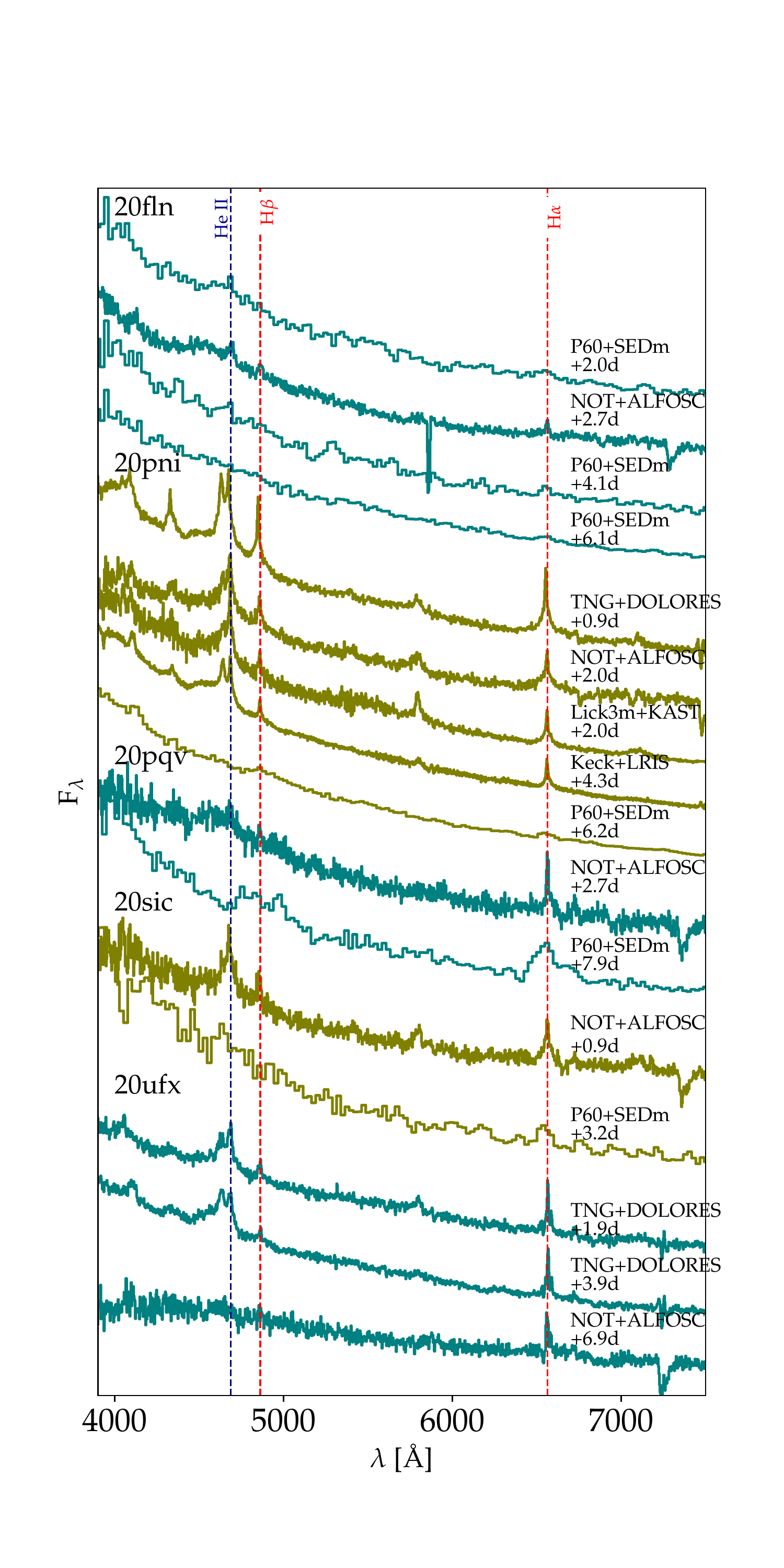}
    \caption{Collection of flash ionisation features spectra for the high-resolution group. The times indicated on the red side of the spectra correspond to the time of acquisition of the spectra from the estimated explosion date.}
    \label{fig:highresflash}
\end{figure*}

\subsection{Subclassification of SNe II}
Following \cite{galyam2017}, we sub-classify our sample in three categories: spectroscopically-normal SNe II, interacting SNe IIn and helium-rich SNe IIb.  Spectroscopically-normal SNe II develop a high velocity ($\approx 10 000$\, km.s$^{-1}$) Balmer P-Cygni profile in the photospheric phase. We do not consider here the photometric subclassifications of SNe IIP or IIL. 
SNe IIn show narrow Balmer emission lines which last for several weeks. SNe IIb develop early on strong absorption lines of He I.\\

\subsection{Flash features in SNe II }

\begin{table*}
\centering
\begin{tabular}{llllll}
\hline
IAU & Type & Flasher & Instrument+Telescope & Time to & Spectrograph \\
name &  &  & (of first spectrum) & first spectrum & Resolution \footnote{Low: $\mathcal{R} < 300$, medium: $300\leq \mathcal{R} \leq 390$, high: $\mathcal{R} > 390$}\\
(SN) &  &  &  & from EED & \\
     &  &  &  & [days] & \\
\hline
\hline
2018grf & SN II & yes & SEDm+P60 & 0.14 & low \\
2019nvm & SN II & yes & SEDm+P60 & 0.17 & low \\
2018dfi & SN IIb & yes & DBSP+P200 & 0.60 & high\\
2020pni & SN II & yes & DOLORES+TNG & 0.86 & high\\
2020sic & SN II & yes & ALFOSC+NOT & 0.89 & high \\
2018dfc & SN II & yes & SEDm+P60 & 1.02 & low \\
2018fif & SN II & yes & DBSP+P200 & 1.13 & high\\
2019ehk & SN IIb & yes & KAST+Lick & 1.47 & high\\
2018cyg & SN II & yes & ACAM+WHT & 1.68 & high\\
2020afdi & SN II & yes & DOLORES+TNG & 1.69 & high \\
2018egh & SN II & yes & ISIS+WHT & 1.86 & high \\
2019ust & SN II & yes & GMOS+Gemini & 1.99 & high \\
\hline
2020lfn & SN II & yes & SEDm+P60 & 2.01 & low\\
2019gmh & SN II & yes & SEDm+P60 & 2.17 & low \\
2020uhf & SN II & yes & SPRAT+LT & 2.21 & medium\\
2018cug & SN II & yes & SEDm+P60 & 2.22 & low \\
2020ufx & SN II & yes & DOLORES+TNG & 2.38 & high \\
2020dcs & SN IIn & yes & SPRAT+LT & 2.48 & medium\\
2018gts & SN II & yes & SEDm+P60 & 2.88 & low \\
2020pqv & SN II & yes & SEDm+P60 & 2.99 & low\\
2018leh & SN II & yes & ISIS+WHT & 3.04 & high\\
2020uqx & SN II & yes & SEDm+P60 & 3.14 & low\\
2020wol & SN II & yes & SEDm+P60 & 3.24 & low\\
2019tjt & SN II & yes & FORS2+VLT & 5.37 & high\\
2019qch & SN II & yes & SEDm+P60 & 6.26 & low\\
2019mor & SN II & yes & SEDm+P60 & 7.44 & low\\
2019lkw & SN II & yes & SEDm+P60 & 9.54 & low\\
2019mge & SN II & yes & SEDm+P60 & 11.40 & low\\
\hline
\end{tabular}
\label{tab:fullflashtable}
\caption{Sample of infant hydrogen rich objects which showed flash ionisation features. They are ordered according to the time from the estimated explosion date until the acquisition of the first spectrum. Events above the horizontal line are those for which a first spectrum was obtained within $\leq 2$ days from EED. }
\end{table*}

We identify flash features using the methods developed in \cite{bruch2021}: we base our identification on the presence of narrow He II emission lines at $\lambda = 4686 $Å.
We identify 28 candidates with flash-ionisation features in our sample (see Table \ref{tab:fullflashtable}). \\
Twelve objects with flash ionisation features had their first spectrum within less than two days from the estimated explosion date. Two of those were classified as SN IIb. Sixteen candidates had their first spectrum more than two days from the EED. Examples of flash ionisation spectra can be found in Figures \ref{fig:lowresflash} and \ref{fig:highresflash}. Examples of spectra not showing flash ionisation can be found in Figure \ref{fig:noflash}.

We describe our flasher candidates in three subgroups: the low-resolution group, for which we obtained spectra mainly with SEDm; the high-resolution group for which we have one or more spectra from higher resolution spectrographs; and the group of objects which show a broad emission line, as described in section 3.3.3 from \cite{bruch2021}.

\subsubsection{The low-resolution group}
This group is composed of  SNe 2018cug, 2018grf, 2019qch, 2019mor, 2019lkw, 2019uqx, 2018gts, 2019mge, 2020uhf, 2020dcs, 2020wol. Those events had their first spectrum taken with either SEDm+P60 or SPRAT+LT, i.e. a resolution lower than 350.
They are displayed in Figure \ref{fig:lowresflash}.  We obtained sequences of SNe 2019qch, 2019mor, 2019lkw and 2019uqx until the He II line was not visible anymore. SNe 2019qch and 2019lkw display the longest lasting flash-ionisation features, respectively fifteen days and nineteen days until the recorded full disappearance of the He II emission line. SNe 20uhf and 20dcs were observed with SPRAT+LT ($R\leq 350$).

\subsubsection{The high-resolution group}
This group is composed of SNe 2018dfc, 2018fif, 2018dfi, 2018leh, 2019tjt, 2019nvm, 2019ehk, 2020fln, 2020pni, 2020pqv, 2020sic and 2020ufx. 
They were followed up with higher-resolution spectrographs ($R>350$). SN2018fif has a short flashing timescale, with He II disappearing within less than 3 days from the estimated explosion time; it was thoroughly studied in \cite{soumagnac2019}. SNe 2018dfi, 2018ehk, 2019nvm also have short timescale of less than 5 days. SNe 2019ehk and 2019tjt show weak flash ionisation features, and would have been counted as non-flashers, if we had not obtained spectra with KAST ($R \approx 500 $) 
 and X-Shooter ($R \approx 5000$), respectively. SN2020pni (see \citealt{terreran2022} and Zimmerman et al. (in prep.)) and 2020ufx both have median flash timescales (around five days) but display very strong emission lines around 1 day after explosion. In both cases we distinguish the He II and N III emission lines. 

\subsubsection{The broad emission feature, aftermath of narrow flash emission}

\begin{figure}
    \centering
    \includegraphics[trim = 20 50 20 75,clip,width=\columnwidth]{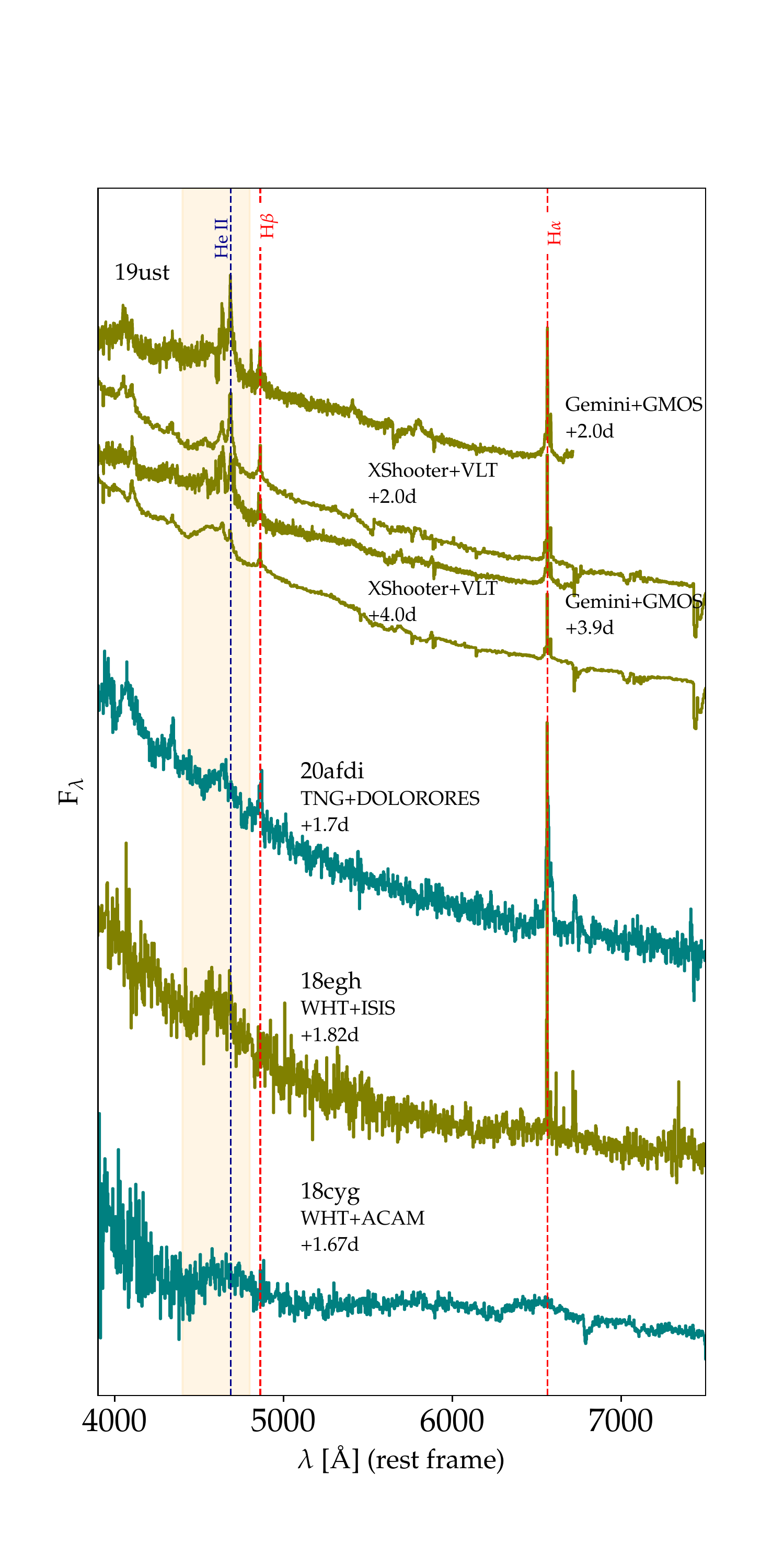}
    \caption{Events showing a broad emission feature (or "bump")}
    \label{fig:bumperflash}
\end{figure}

SNe 2018cyg and 2018egh  were previously classified as dubious flashers due to presence of a broad feature in lieu of a narrow He II emission lines. SN 2020afdi also shows a broad feature rather than narrow emission lines at early time.  However, we observed in SN 2019ust the transition from narrow He II lines to a broad structure within four days from estimated explosion time, see Figure \ref{fig:bumperflash}. We hence assume that such a broad feature at early time is a signature of flash ionisation features. Such broad feature could be the result of the blending of a forest of lines. To our knowledge, there is no investigation of this structure in the literature. In 2019ust time sequence, this broad feature seems to originate from the blending of fading lines such as He II and N III. 

 In 19ust, this broad structure marks a transitional phase when the flash feature phase ends. The structure itself disappears afterwards. Hence, we now include events with this broad feature as identified flashers. SNe 2020adfi, 2018cyg and 2018egh have hence short flash-feature timescale (less than two days from EED).

It is clear however that low-resolution and low-throughput spectrographs could easily miss either low contrast He II lines or the broad emission structure around 4800 Å.






\section{Results and discussion} \label{resultsdiscuss}

\subsection{Fraction of SNe II with flash ionisation features}

As in \cite{bruch2021}, we estimate the fraction of SNe II with flash ionisation features using candidates who had a first spectrum within less than two days from EED. Twenty four spectroscopically-normal SNe II had a first spectrum within less than two days from EED, ten show flash ionisation emission lines, while fourteen did not.

\begin{table}
\centering
\hspace{-2cm}
\renewcommand{\arraystretch}{0.82}
\begin{tabular}{lllll}
\hline
IAU & Flasher & Time to FS & App. Mag & SNR \\
name &  & from EED & at FS &  \\
(SN) &  & [days] & [AB mag] &  \\
\hline
\hline
2019ewb & no & 1.08 & 19.60 & 3.15 \\
\hline
2020sjv & no & 1.51 & 18.55 & 5.01 \\
2019ikb & no & 1.94 & 17.44 & 5.06 \\
2020xhs & no & 1.89 & 18.73 & 5.39 \\
2019omp & no & 1.05 & 19.33 & 5.43 \\
2020acbm & no & 0.17 & 18.11 & 11.90 \\
2020dyu & no & 1.12 & 18.69 & 12.56 \\
\hline
2020dya & no & 1.38 & 18.67 & 15.47 \\
2020abbo & no & 1.21 & 18.46 & 15.52 \\
2018iuq & no & 0.10 & 17.77 & 20.04 \\
2019nvm & yes & 0.17 & 17.78 & 22.83 \\
2018grf & yes & 0.14 & 18.91 & 23.53 \\
2018dfc & yes & 1.02 & 18.04 & 24.09 \\
\hline
\end{tabular}
\caption{SNR estimation of the first spectra obtained with SEDm in the 2-day subsample.}
\label{tab:sedm2d}
\end{table}

\begin{figure}
   \centering
    \includegraphics[trim={2.5cm 0 1cm 1cm},clip,width = 1.1\columnwidth]{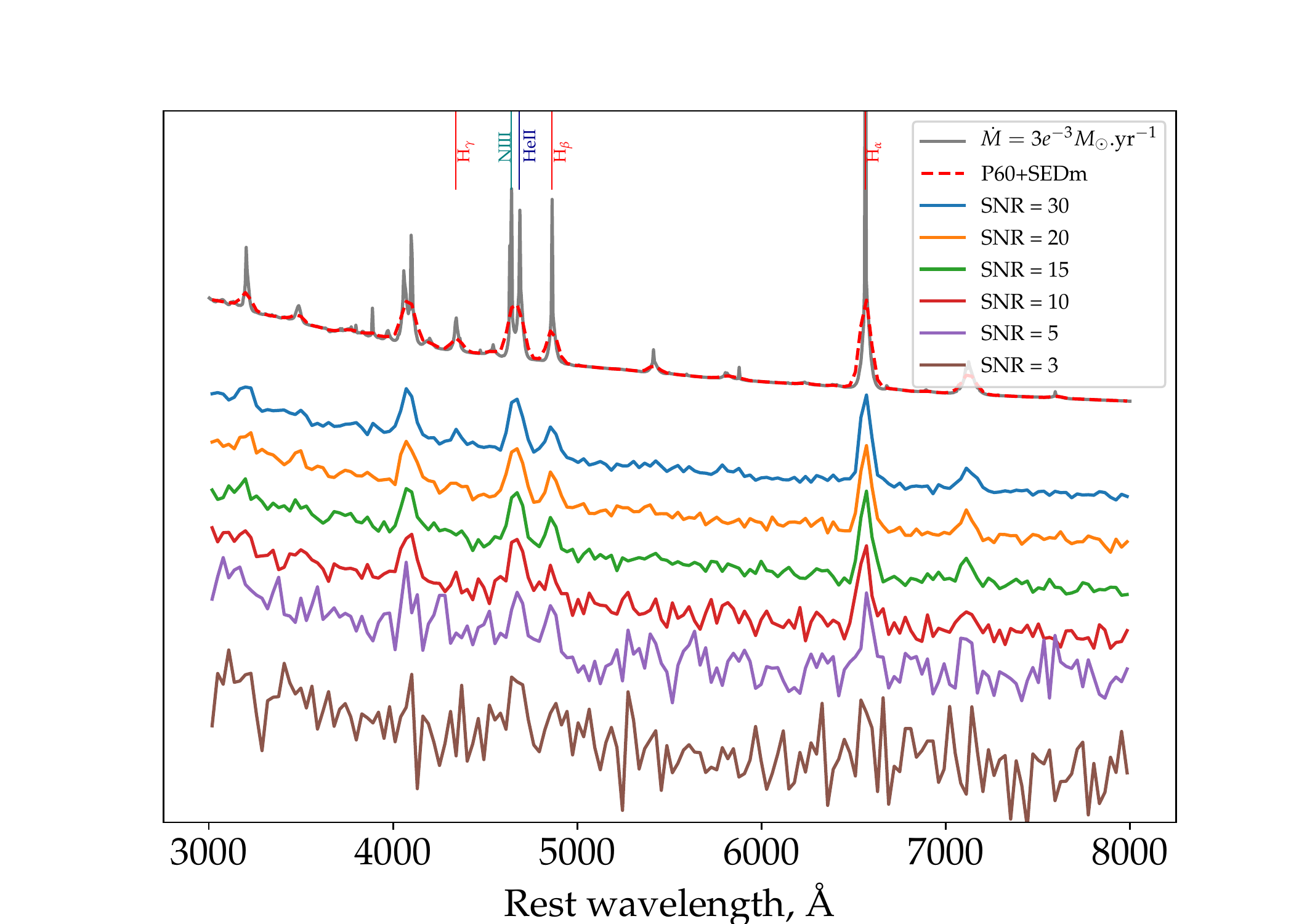}\\
    \includegraphics[trim={2.5cm 0 1cm 1cm},clip,width = 1.1\columnwidth]{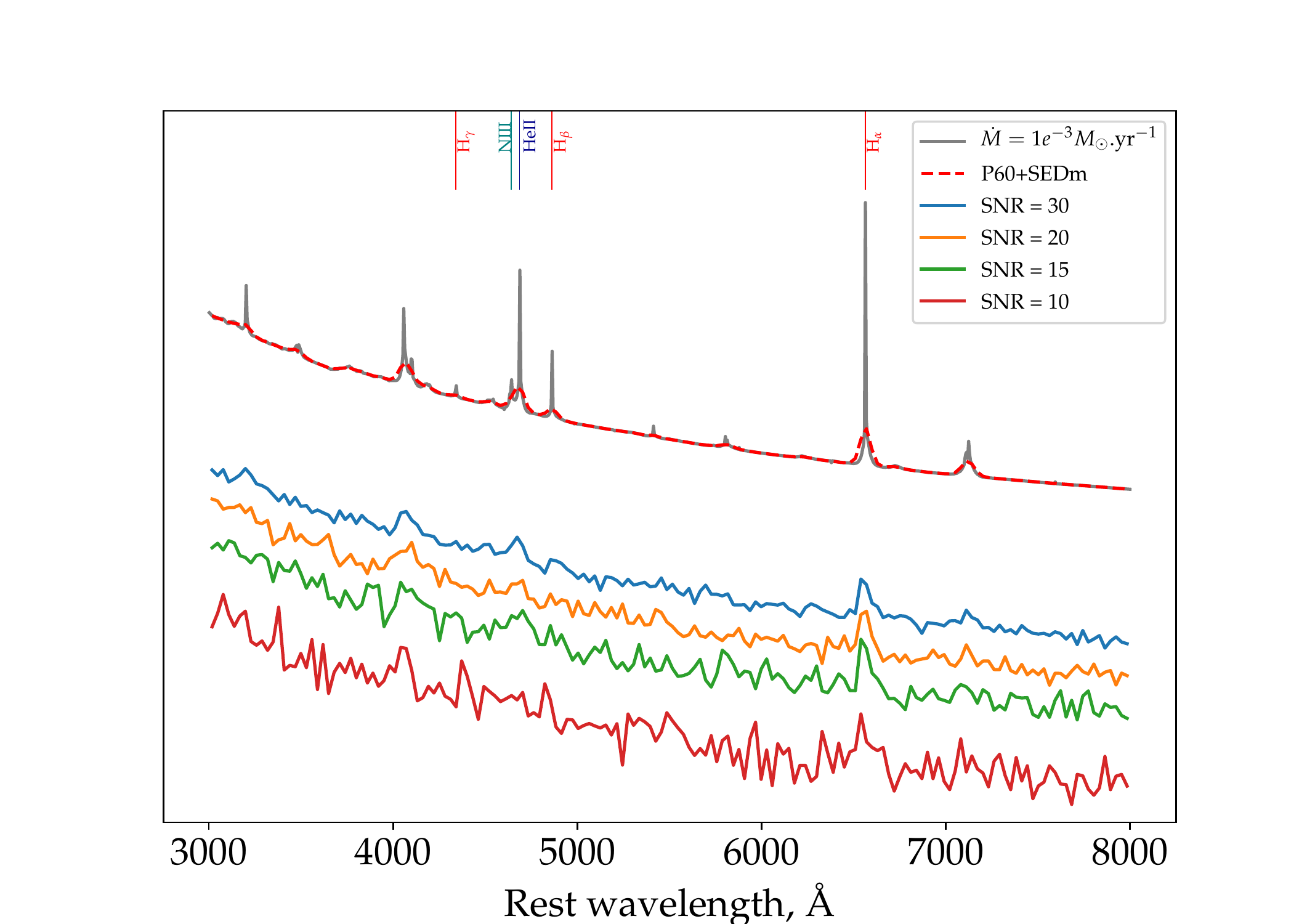}
    \caption{Simulating noisy SEDm spectra. Top: the strong flash-ionisation features template is used. With a SNR below 5, it is hard to detect even strong flash-ionisation features. Bottom: Weak flash ionisation template. Weak flash features can only be detected for a SNR above 15.}
    \label{fig:SNR_sedm}
\end{figure}

\begin{figure}
    \centering
    \includegraphics[trim={2.5cm 0 1cm 1cm},clip,width = 1.1\columnwidth]{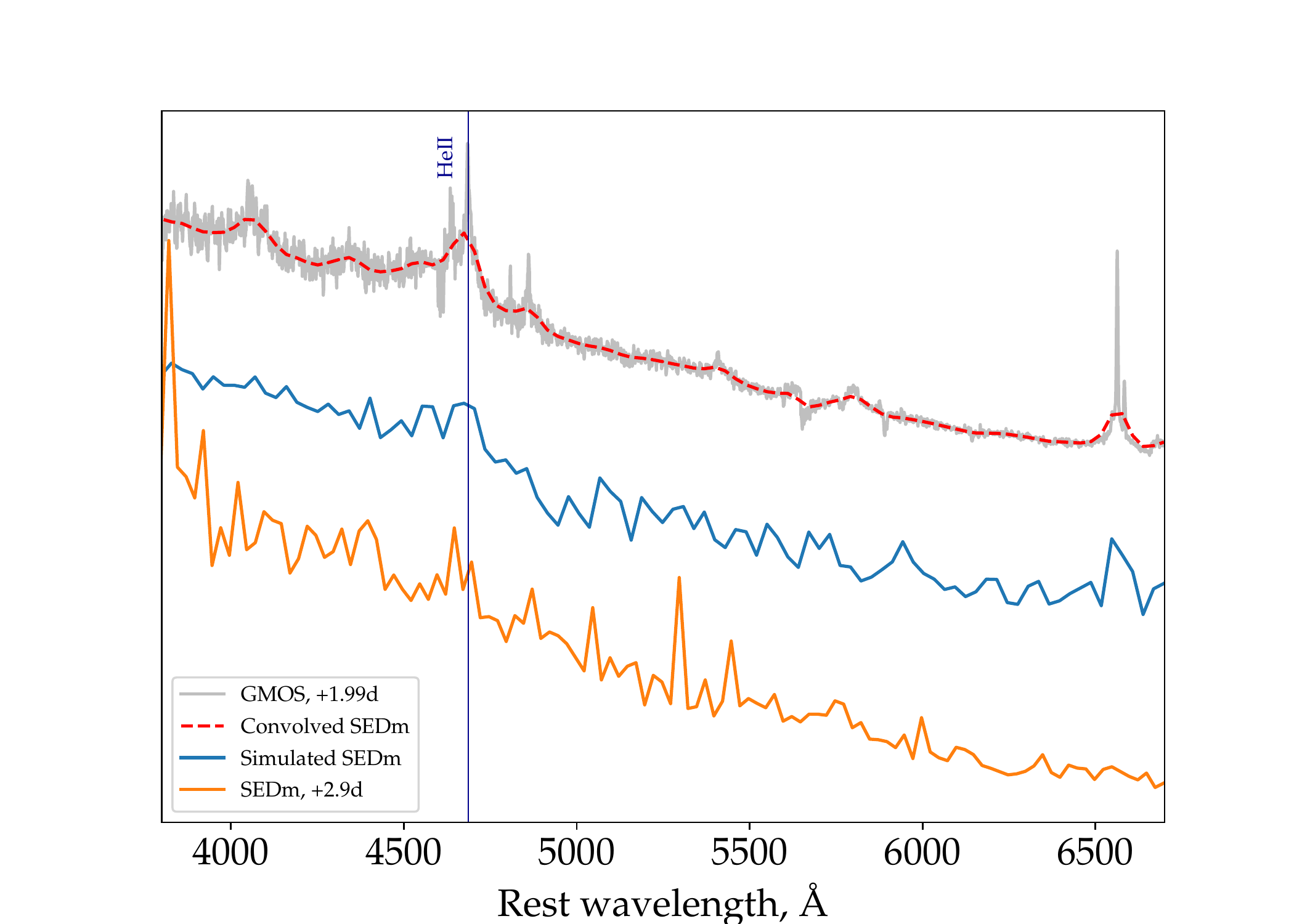}
    \caption{Simulation of an SEDm spectrum from a high-resolution GMOS spectrum, 2d after the estimated explosion. In blue is the simulated SEDm spectrum and in orange is a real SEDm spectrum obtained 3 days after estimated explosion time. Flash-ionisation features are visible in the GMOS spectrum. Once convolved to the SEDm resolution and noised, the flash features are indistinguishable from the noise. Without the GMOS spectrum, this candidate would have been classified as a non-flasher. }
    \label{fig:19ust}
\end{figure}

\begin{figure}[h!]
\hspace{-0.75cm}
    \includegraphics[trim={1cm 0 1cm 1cm},clip,width = 1.1\columnwidth]{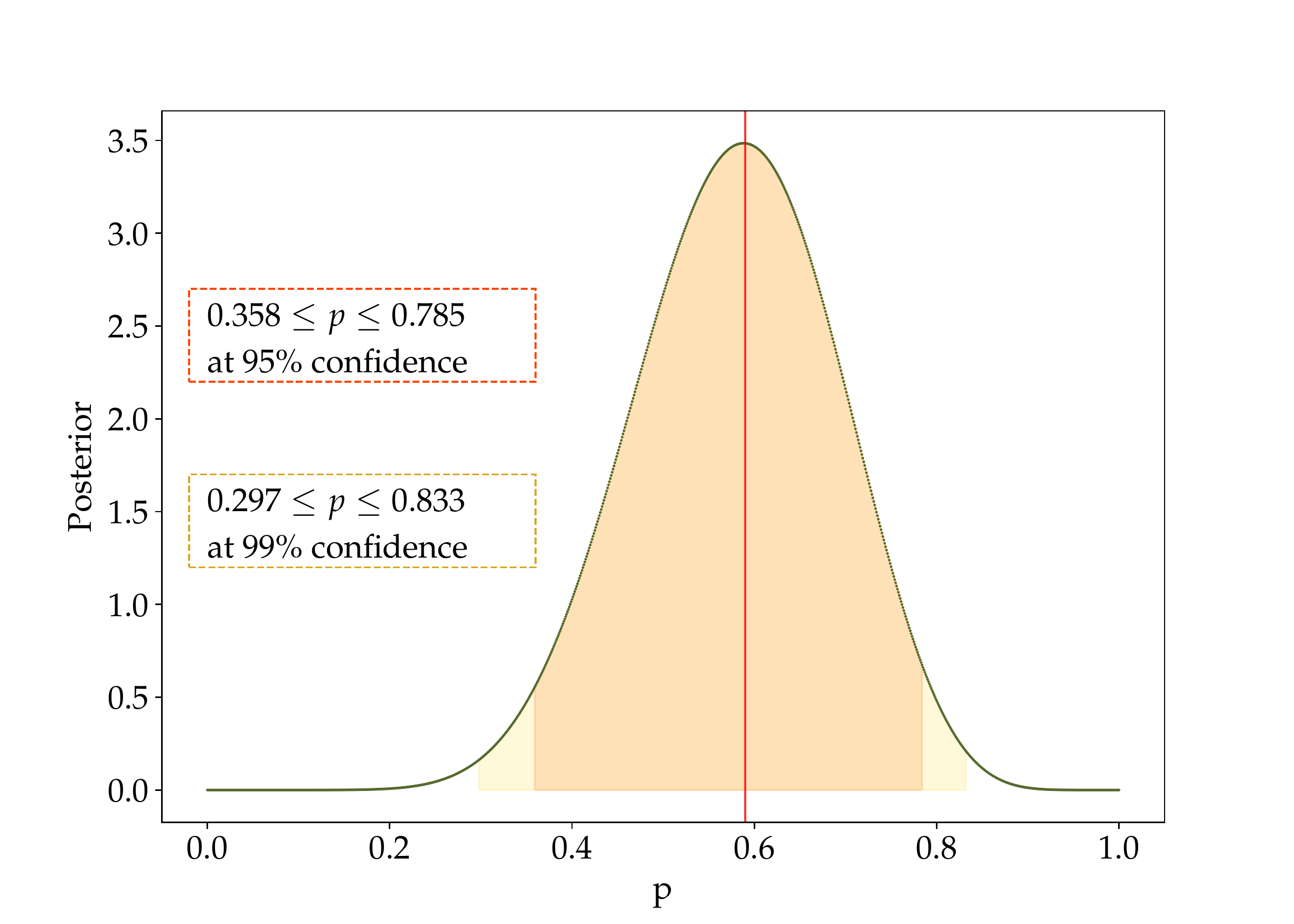}
    \caption{Posterior distribution of the probability for observing flash-features in the sample of candidates who had a first spectrum within less than 2 days from EED, having rejected candidates with a an SNR lower than 15 in their SEDm spectrum.}
    \label{fig:strongweakflash}
\end{figure}

We used SEDm to obtain early spectra mainly due to its high availability and short response time. Being co-located with the survey telescope, we can obtain a spectrum $<1$ hour after we submit a trigger. However, due to the relatively small diameter of the telescope used (60"), the overall sensitivity of SEDm is limited. This makes it challenging to detect flash features in some spectra. For example, in Figure \ref{fig:19ust}, we show a high resolution spectrum obtained with GMOS ($\mathcal{R}>1500$) and degrade it to the resolution of SEDm. Next, we change the SNR of the degraded spectrum to the measured SNR of a nearly contemporaneous SEDm spectrum. This shows that we cannot distinguish the flash lines (i.e. He II, and Balmer lines H$\alpha$ and H$\beta$) from the noisy continuum anymore. Hence if we had relied on a SEDm spectrum to identify flash features, this candidate would have been mislabeled as a non-flasher. \\ 
We examine the SNR of the ten non-flashers whose first spectrum was obtained with SEDm, see Table \ref{tab:sedm2d}. In order to estimate a SNR threshold below which we cannot discriminate between flashers and non-flashers, we use the flash-spectra templates by \cite{grohb2020} and degrade them to the resolution of SEDm and inject noise. We use templates with high mass-loss rate ($\dot{M} \approx 3\times10^{-3}\,M_{\odot}$.yr$^{-1}$) to simulate strong flashers, and lower mass-loss rates ($\dot{M}\approx 1\times 10^{-3}\,M_{\odot}$.yr$^{-1}$) as weak flashers\footnote{We use the templates with $v_{inf} = 150\,$km.s$^{-1}$, $R_{*}=8.10^{13}$cm and a CNO processed-like surface abundance. They are publicly available templates on WiseRep.}, see Figure \ref{fig:strongweakflash}. For weak flashers, a SNR $< 15$ cannot be used to identify flash features. For strong flashers, a SNR $<5$ is unusable. \\

\begin{figure}
    \centering
    \includegraphics[trim = 20 50 20 75,clip,width=\columnwidth]{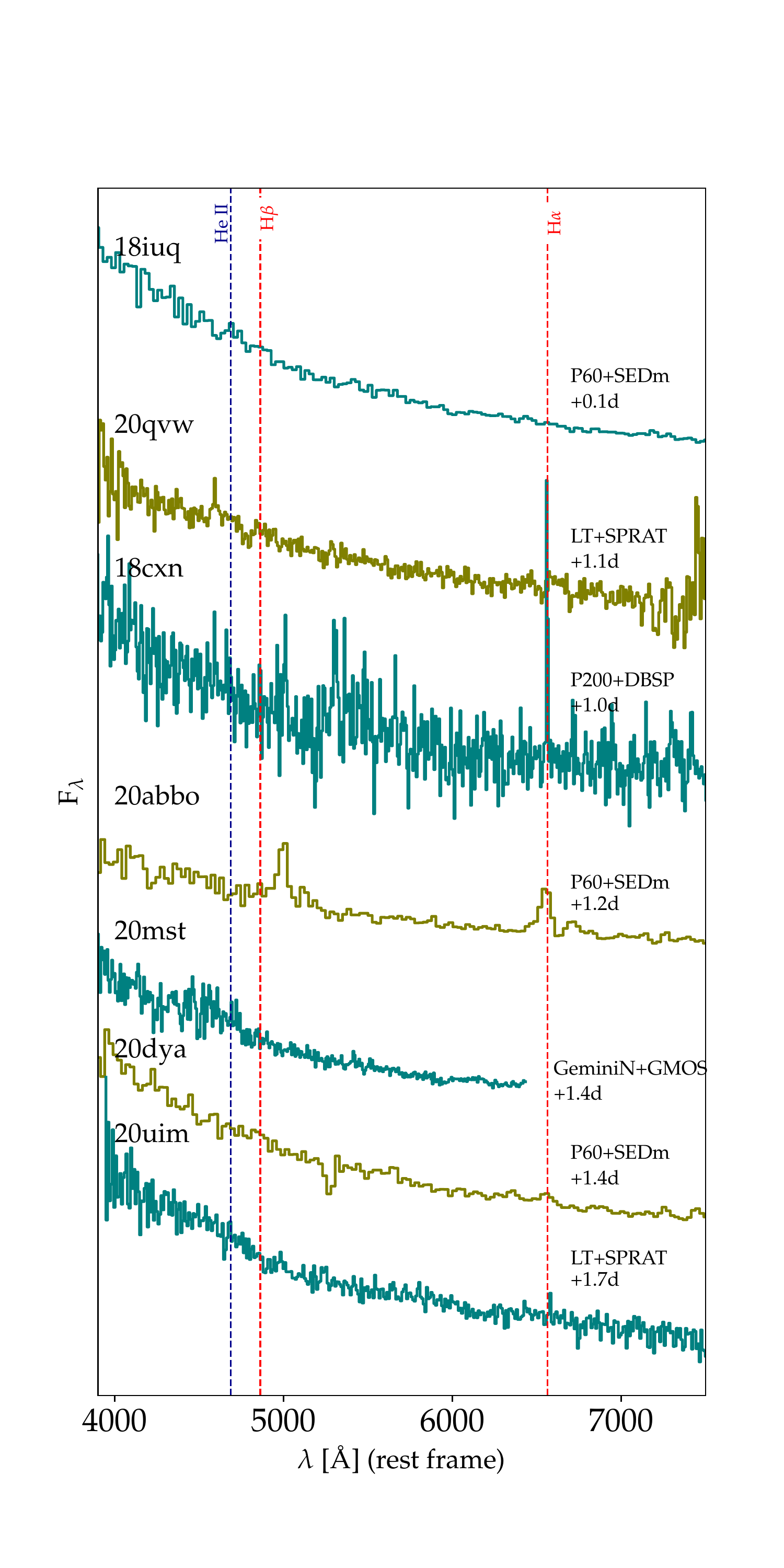}
    \caption{Spectra of the non-flashers from the 2-day subsample, excluding those which has a SNR lower than 15.}
    \label{fig:noflash}
\end{figure}

We eliminate seven events who have a SNR lower than 15,which leaves us with seventeen candidates, see Table \ref{tab:weaksubsamp2d}. The spectra of the non flashers remaining in this sample can be found in Figure \ref{fig:noflash}.
The fraction of objects with CSM prior to explosion is then $58.8\% ^{+19.7}_{-23}$, at 95\% confidence interval (CI). In the unlikely case where we consider all flashers are strong flashers, only one candidate has a SNR lower than 5. In this case, the fraction lowers to $43.5\% ^{+19.8}_{-17.9}$, at 95\% CI  (10 out of 23 have show flash features), see Figure \ref{fig:strongweakflash}. We conclude that it is likely that most progenitors of SNe II are embedded in CSM. These new results are consistent with our previously estimated fraction, \cite{bruch2021}. \\

\subsection{Comparison of the photometric properties of flashers and non-flasher events}

\begin{table*}
\caption{Subsample of SN II objects with a first spectrum within $< 2$\,d from EED. Objects with spectra whose SNR are below 15 were removed}
\hspace{-3cm}
\begin{tabular}{lllllllllll}
\hline
IAU & Flasher & Time to  & Telescope +  & Redshift & Error on & Band & Peak Absolute & Error on & Rise & Error on  \\
name &  & first spectrum  & Instrument  &  & redshift &  & magnitude & peak absolute & time & rise time  \\
 &  &   &  &  &  &  &  & magnitude &  &   \\
(SN) &  & [d]  &  &  &  &  & [AB mag] & [AB mag] & [d] & [d]  \\
\hline
\hline
2018iuq & no & 0.11 & SEDm+P60 & 0.026 & 0.0047 & r & -18.60 & 0.39 & 12.18 & 1.30  \\
 & & & & & & g & -18.75 & 0.39 & 11.51 & 0.39 \\
2018grf & yes & 0.14 & SEDm+P60 & 0.054 & 0.0073 & r & -18.52 & 0.30 & 6.29 & 0.39  \\
 & & & & & & g & -18.58 & 0.30 & 4.88 & 0.29 \\
2019nvm & yes & 0.17 & SEDm+P60 & 0.018 & $<10^{-4}$ & r & -17.64 & 0.01 & 8.58 & 0.57 \\
 & & & & & & g & -17.47 & 0.02 & 7.68 & 1.22 \\
2020qvw & no & 0.71 & SPRAT+LT & 0.055 & 0.0028 & r & -18.40 & 0.13 & 10.83 & 0.59 \\
& & & & & & g & - & - & - & - \\
2020pni & yes & 0.86 & DOLORES+TNG & 0.017 & $<10^{-4}$ & r&-18.25 & 0.01 & 11.01 & 0.27 \\
& & & & & & g & -18.27 & $<10^{-2}$ & 6.27 & 0.53 \\
2020sic & yes & 0.89 & ALFOSC+NOT & 0.033 & 0.0001 & r&-17.87 & 0.29 & 12.37 & 1.57 \\
& & & & & & g & - & - & - & - \\
2018cxn & no & 0.99 & DBSP+P200 & 0.041 & 0.0001 & r&-17.49 & 0.01 & 16.19 & 0.29 \\
& & & & & & g & -17.51 & 0.01 & 9.52 & 0.62 \\
2018dfc & yes & 1.02 & SEDm+P60 & 0.037 & 0.0001 & r &-18.50 & 0.02 & 10.43 & 1.50 \\ 
& & & & & & g & -18.55 & 0.01 & 7.37 & 1.35 \\
2018fif & yes & 1.13 & DBSP+P200 & 0.017 & $<10^{-4}$ & r &-17.18 & 0.01 & 14.13 & 1.03 \\ 
& & & & & & g & -17.02 & 0.03 & 10.04 & 1.69 \\
2020abbo & no & 1.21 & SEDm+P60 & 0.017 & 0.0012 & r &-16.66 & 0.15 & 24.76 & 1.93 \\
& & & & & & g & -16.53 & 0.15 & 11.27 & 0.27 \\
2020mst & no & 1.30 & GMOS+Gemini & 0.058 & 0.0108 & r & -18.13 & 0.41 & 15.06 & 0.72 \\
& & & & & & g & -18.09 & 0.41 & 10.90 & 0.84 \\
2020dya & no & 1.38 & SEDm+P60 & 0.030 & 0.0001 & r & -17.55 & 0.01 & 15.08 & 0.46 \\
& & & & & & g & - &- & - & - \\
2018cyg & yes? & 1.68 & ACAM+WHT & 0.012 & 0.0002 & r &-15.49 & 0.03 & 16.86 & 0.29 \\ 
& & & & & & g & -14.51 & 0.03 & 10.37 & 0.30 \\
2020afdi & yes & 1.69 & DOLORES+TNG & 0.024 & 0.0001 & r & -15.88 & 0.01 & 6.26 & 0.92 \\
& & & & & & g & -15.94 & 0.01 & 4.72 & 0.55 \\
2020uim & no & 1.72 & SPRAT+LT & 0.018 & 0.0001 & r &-16.78 & 0.02 & 18.62 & 1.46 \\
& & & & & & g & -16.87 & 0.02 & 7.19 & 0.73 \\
2018egh & yes? & 1.86 & ISIS+WHT & 0.038 & 0.0001 & r &-16.81 & 0.01 & 16.74 & 0.78 \\
& & & & & & g & -16.66 & 0.03 & 5.72 & 0.60 \\
2019ust & yes & 1.99 & GMOS+Gemini & 0.022 & $<10^{-4}$ & r &-18.17 & 0.02 & 12.47 & 1.65 \\
& & & & & & g & -18.02 & $<10^{-2}$ & 15.15 & 0.11 \\
\hline
\end{tabular}
\label{tab:weaksubsamp2d}
\end{table*}


Strong CSM interaction may provide an additional power source, resulting usually in brighter events such as SNe IIn, see \cite{Smith2016} and \cite{nyholm2020}. Since flash features also arise from CSM interaction, we want to test whether their early light curve behaviour differs from non-flashers. It was suggested that events showing flash features at early time would be brighter, see \cite{Hosseinzadeh2018}. \\
In order to test this claim, we measure the peak absolute magnitude and rise time, as well as the color at peak in $g$ band of our sample, using the methods described earlier. We restrict our test to the subsample of normal SNe II for which we could robustly discriminate between flashers and non flashers, i.e. where we consider weak flashers (17 objects). We call this sub sample as the golden 2-day subsample, see Figure \ref{fig:subsamps} in the Appendix. \\ 

\begin{figure*}
    \centering
    \includegraphics[trim = 0 0 60 10,clip,width= 1\columnwidth]{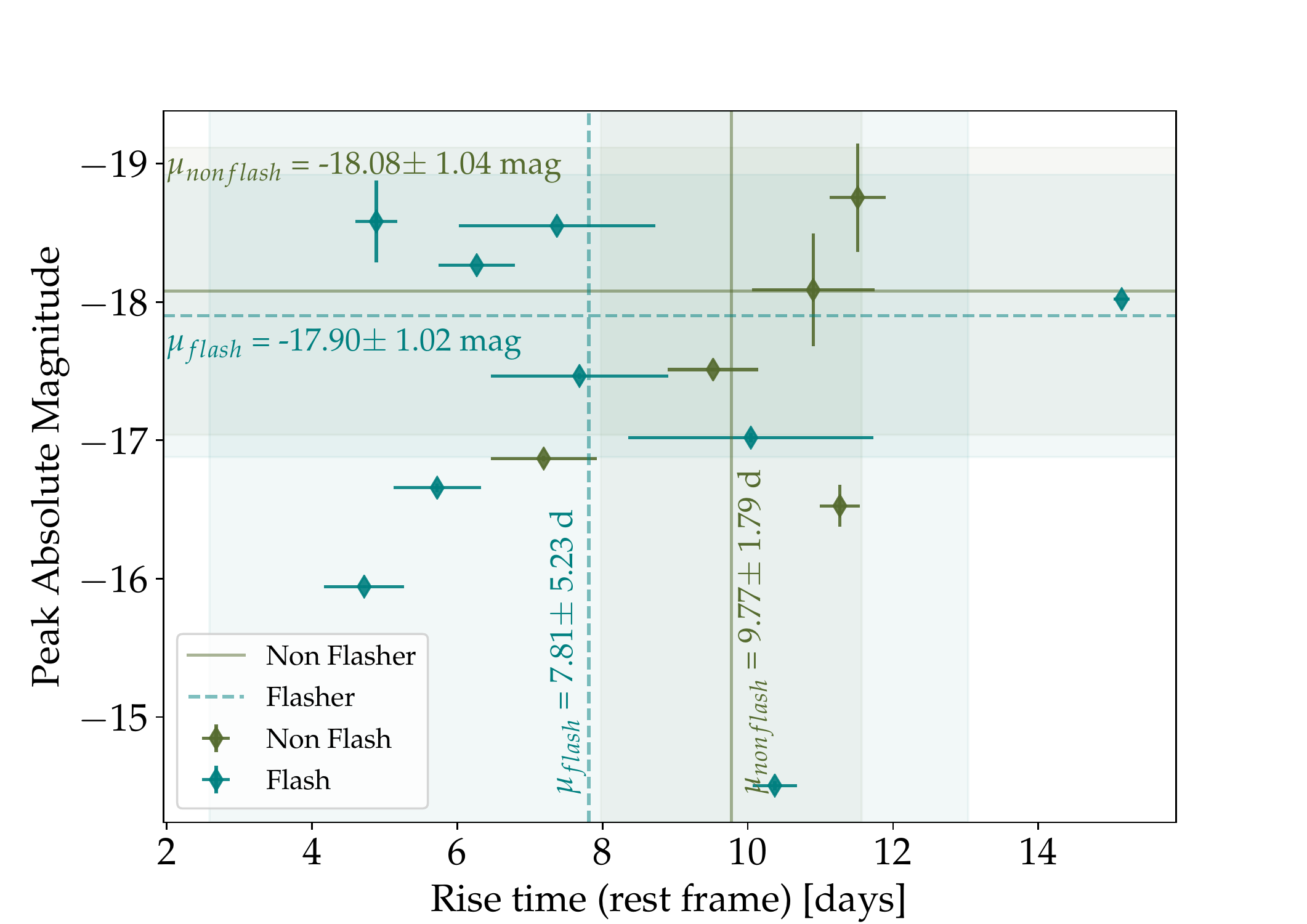}
    \includegraphics[trim = 0 0 60 10,clip,width= 1\columnwidth]{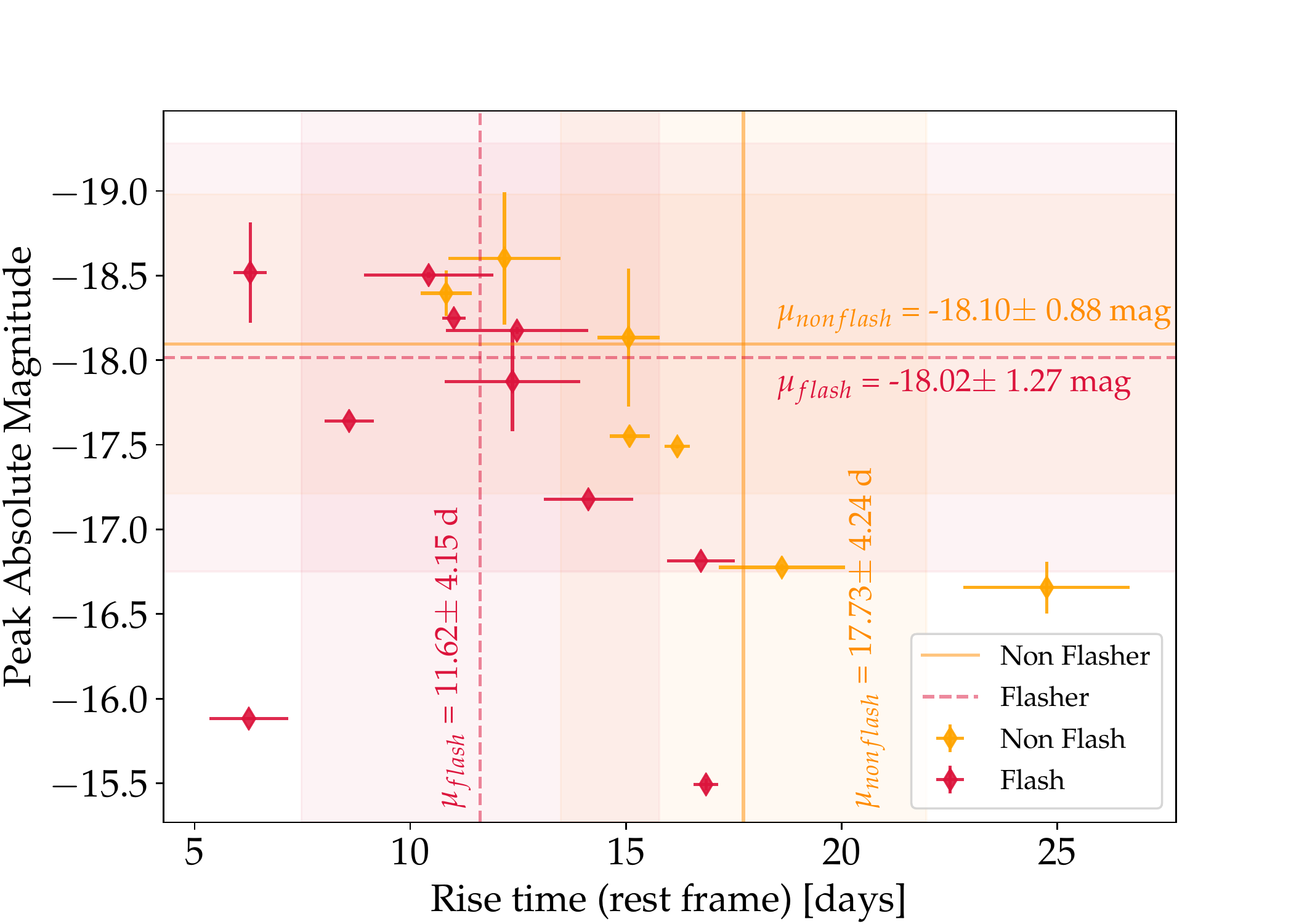}
    \caption{Peak absolute magnitude vs rise time for the 2d subsample in $g$ band (left) and $r$ band (right). The solid dark green (orange) lines indicate the weighted mean of the peak absolute magnitude (horizontal) and the weighted mean of the rise time (vertical) of non-flashers, the dashed teal (red) lines for flashers. The shaded area correspond to the corresponding weighted standard deviation.}
    \label{fig:peakrise}
\end{figure*}

\begin{figure*}[t!]
    \includegraphics[trim = 0 0 50 50,clip,width= 1\columnwidth]{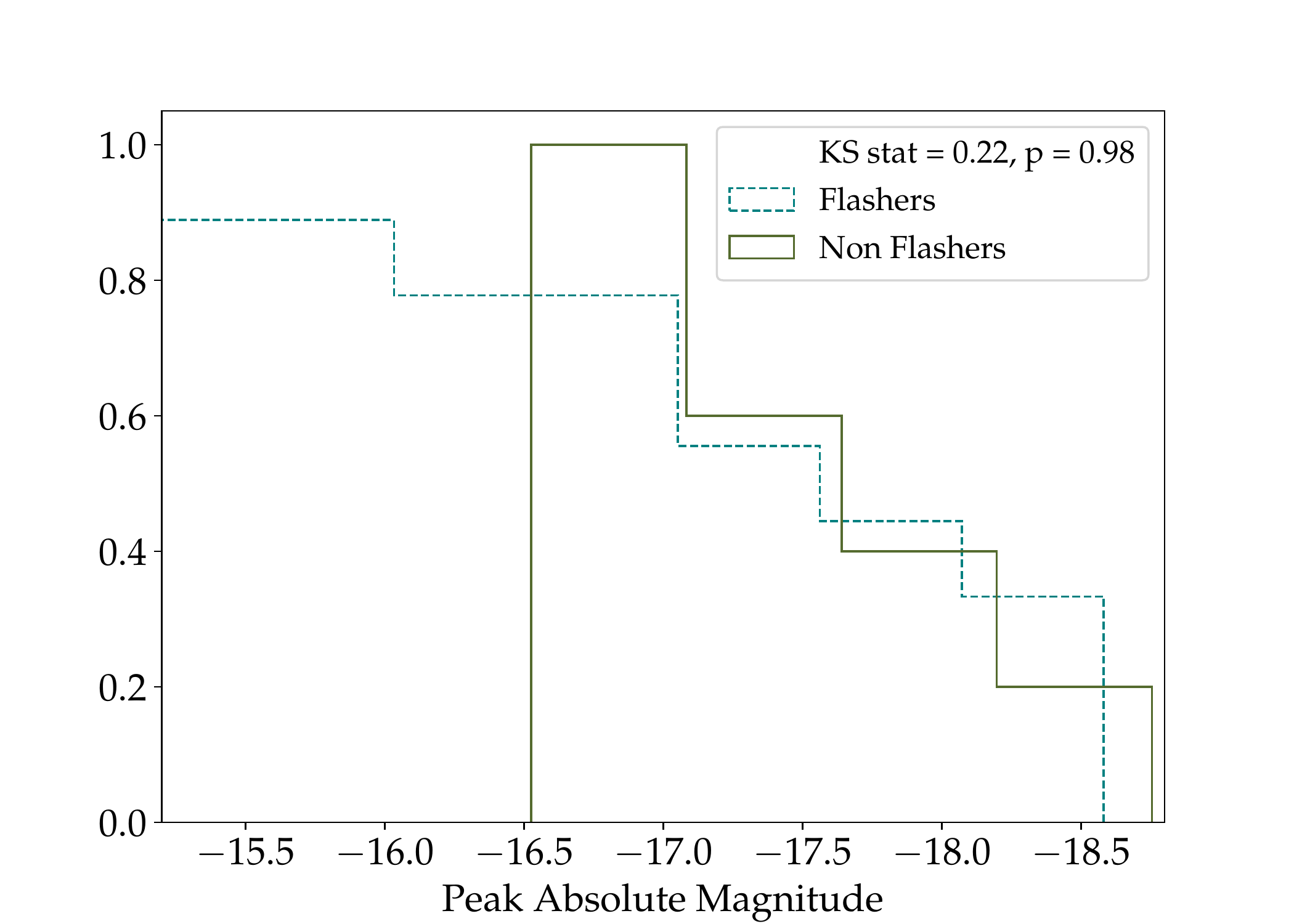}
    \includegraphics[trim = 0 0 50 50,clip,width= 1\columnwidth]{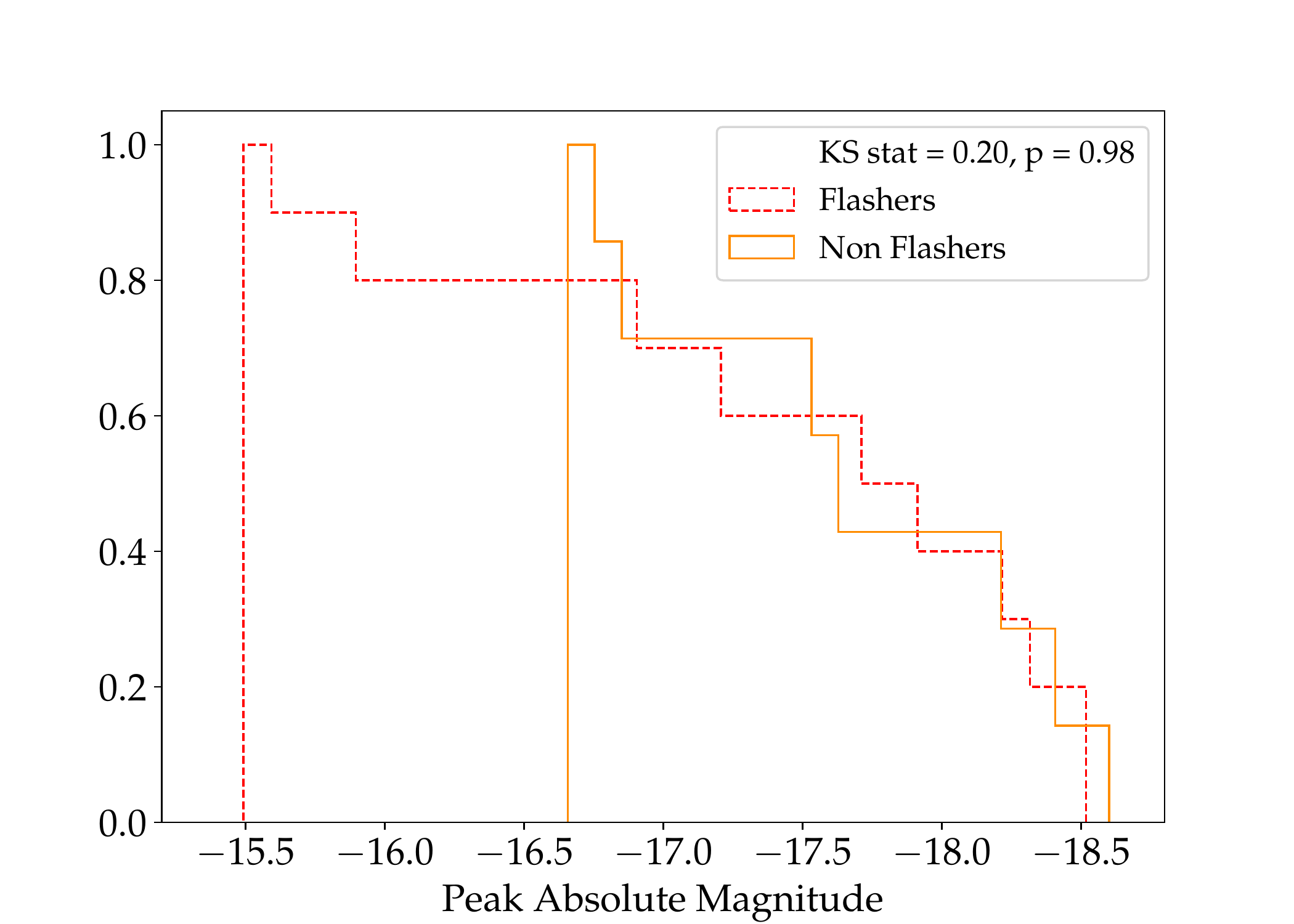}
    \caption{KS tests for the peak absolute magnitude distribution of flashers and non flashers in green (left) and red (right) bands. }
    \label{fig:peakKSTEST}
\end{figure*}

\begin{figure*}
    \includegraphics[trim = 0 0 50 50,clip,width= 1\columnwidth]{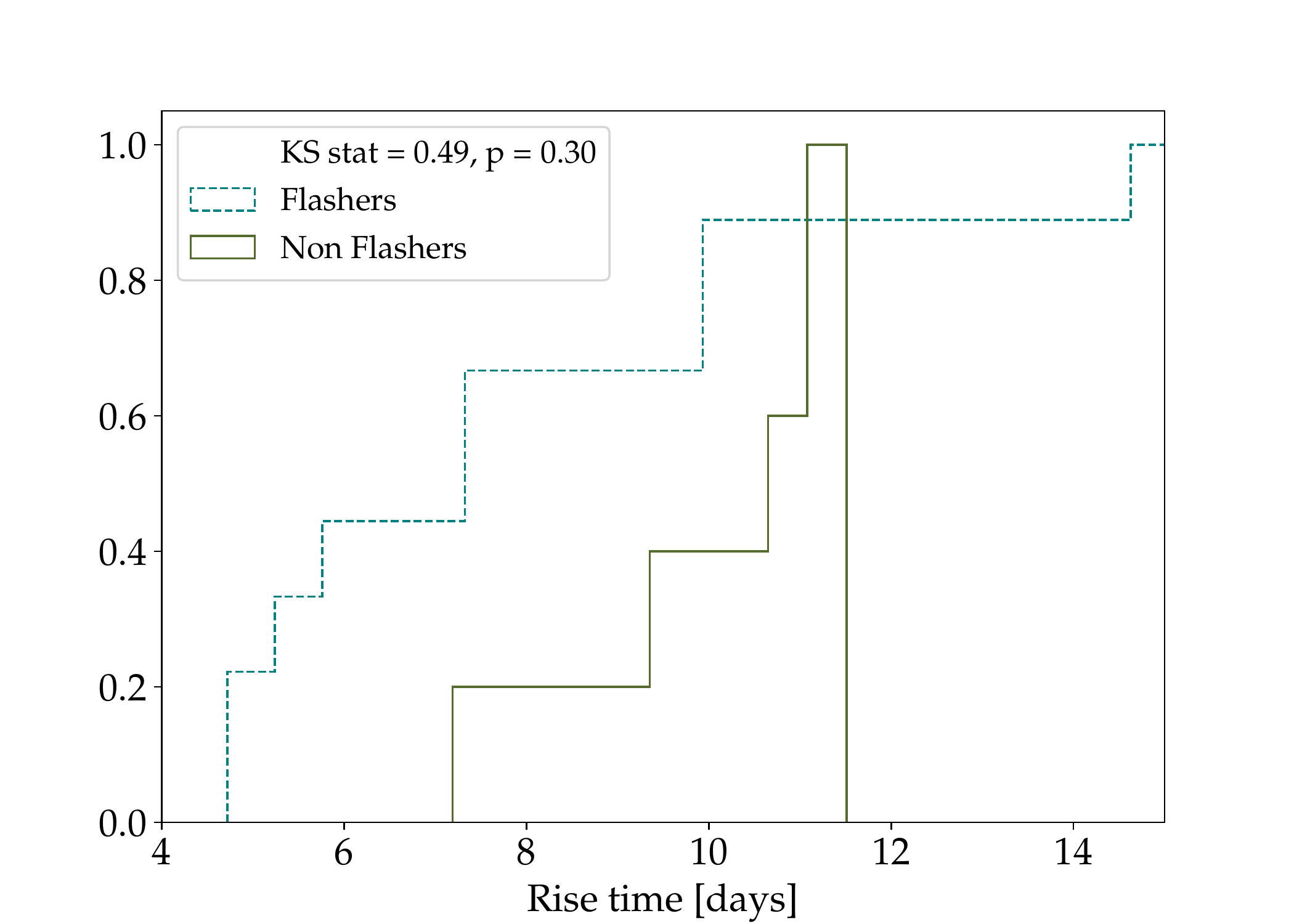}
    \includegraphics[trim = 0 0 50 50,clip,width= 1\columnwidth]{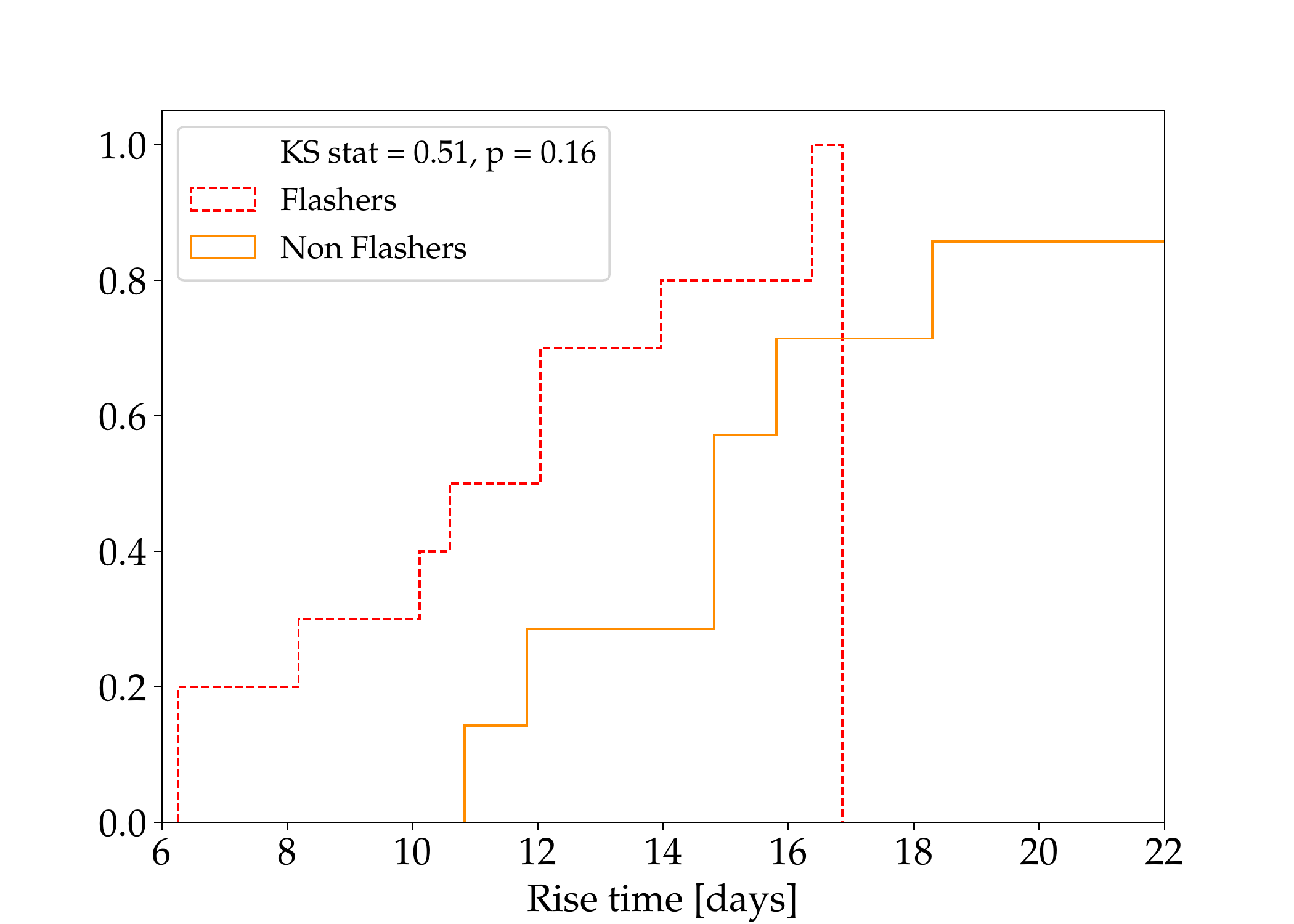}
    \caption{KS tests for the rise time distribution of flashers and non flashers in green (left) and red (right) bands. }
    \label{fig:riseKSTEST}
\end{figure*}

We calculate the weighted mean of the peak absolute magnitude and rise time, and the standard deviation on the weighted mean. We find that flashers and non-flashers have almost identical peak magnitude distribution, see Fig. \ref{fig:peakrise}. Their mean values, M$_{flash} = -18.02 \pm 1.27$ mag and M$_{noflash} = -18.10 \pm 0.88$ mag in the $r$ band and M$_{flash} = -17.90 \pm 1.02$ mag and M$_{noflash} = -18.08 \pm 1.04$ mag in the $g$ band show that flashers are not brighter than non-flashers. A Kolmogorov-Smirnoff (KS) test reveals that the absolute peak magnitudes of flashers and non-flashers are not significantly different ($p_{value} = 0.98$) in either of the bands (see Figure \ref{fig:peakKSTEST}).\\
Flashers and non-flashers also have similar rise times (Figure \ref{fig:peakrise}), with flashers rising to peak in t$^g_{rise} = 7.81 \pm 5.23$ days in the $g$ band and t$^r_{rise} = 11.62 \pm 4.15$ days in the $r$ band compared to non-flashers : t$^g_{rise} = 9.77\pm 1.79$ days in the $g$ band and t$^r_{rise} = 17.73 \pm 4.24$ days in the $r$ band. The KS test in the $g$ and $r$ band returns $p_{value,g} = 0.30$ and $p_{value,r} = 0.16$, respectively. As these values are higher than the threshold for a significant detection ($p=0.05$), we cannot reject the null hypothesis that these two distributions are drawn from the same parent distribution (see Figure \ref{fig:riseKSTEST}).

We also investigate the color at the $g$ band peak . We use the interpolated light curves and subtract theirs values at peak g band.
One candidate (SN2018cyg) has significant host extinction ($g-r$ = 0.9 mag). We had previously estimated the host extinction for this event using the method derived by \cite{Poznanski_2012} and found that the absolute peak magnitude in the $g$ band is estimated to be extincted by nearly four magnitudes, see section 4.2 in \cite{bruch2021}. Excluding this object, the distribution in color at peak for flashers and non-flashers is similar. A KS test returns a p-value of 0.69, which indicates that the colors at peak in the $g$ band for flashers and non-flashers are also drawn from the same parent distribution.

At early time, SNe II with flash-ionisation features behave similarly to those without. This indicates that the CSM creating the flash features is not massive enough to contribute significantly to the luminosity of SNe II.

\begin{figure}[h!]
    \includegraphics[width= 1.1\columnwidth]{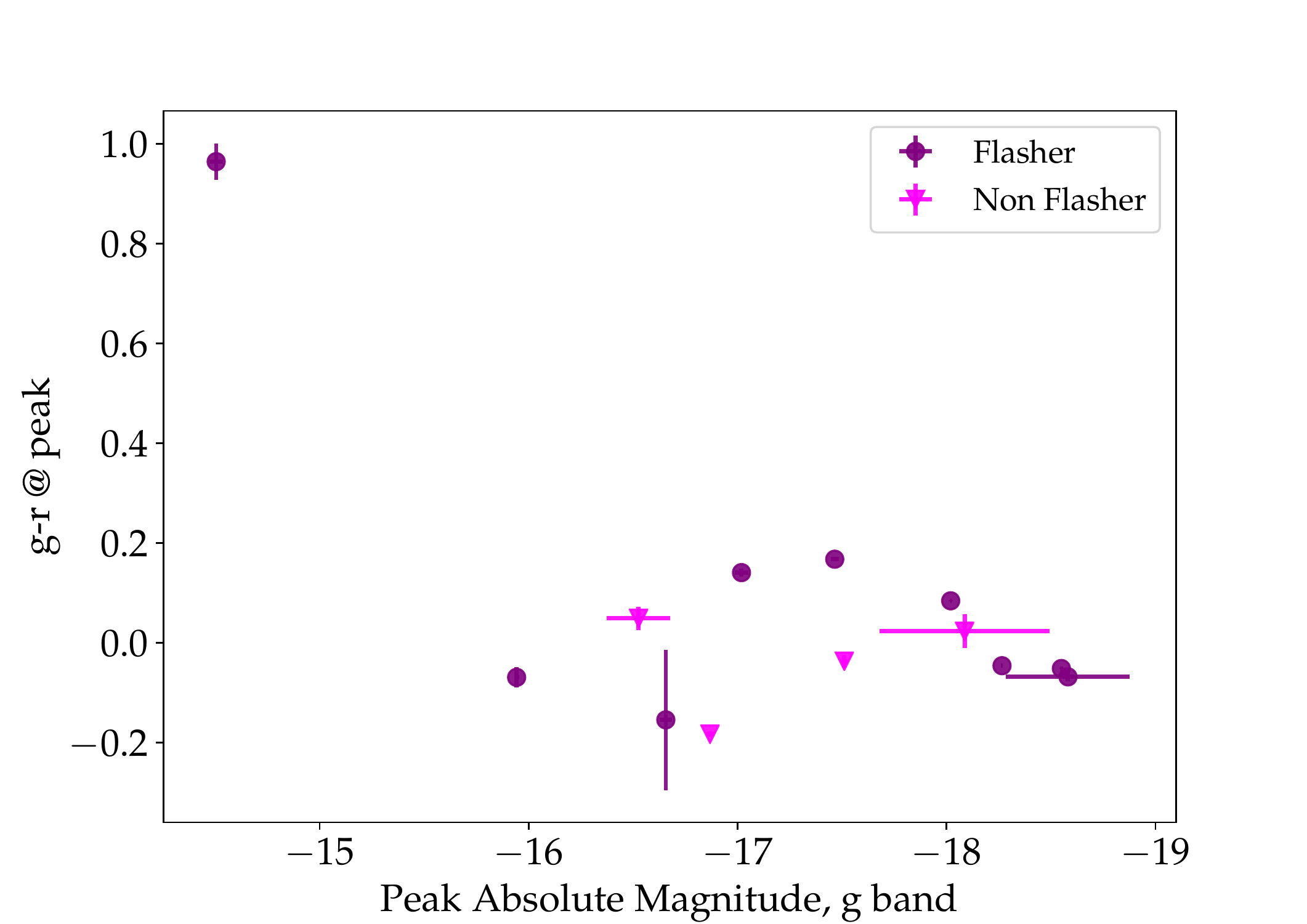}
    \includegraphics[width= 1.1\columnwidth]{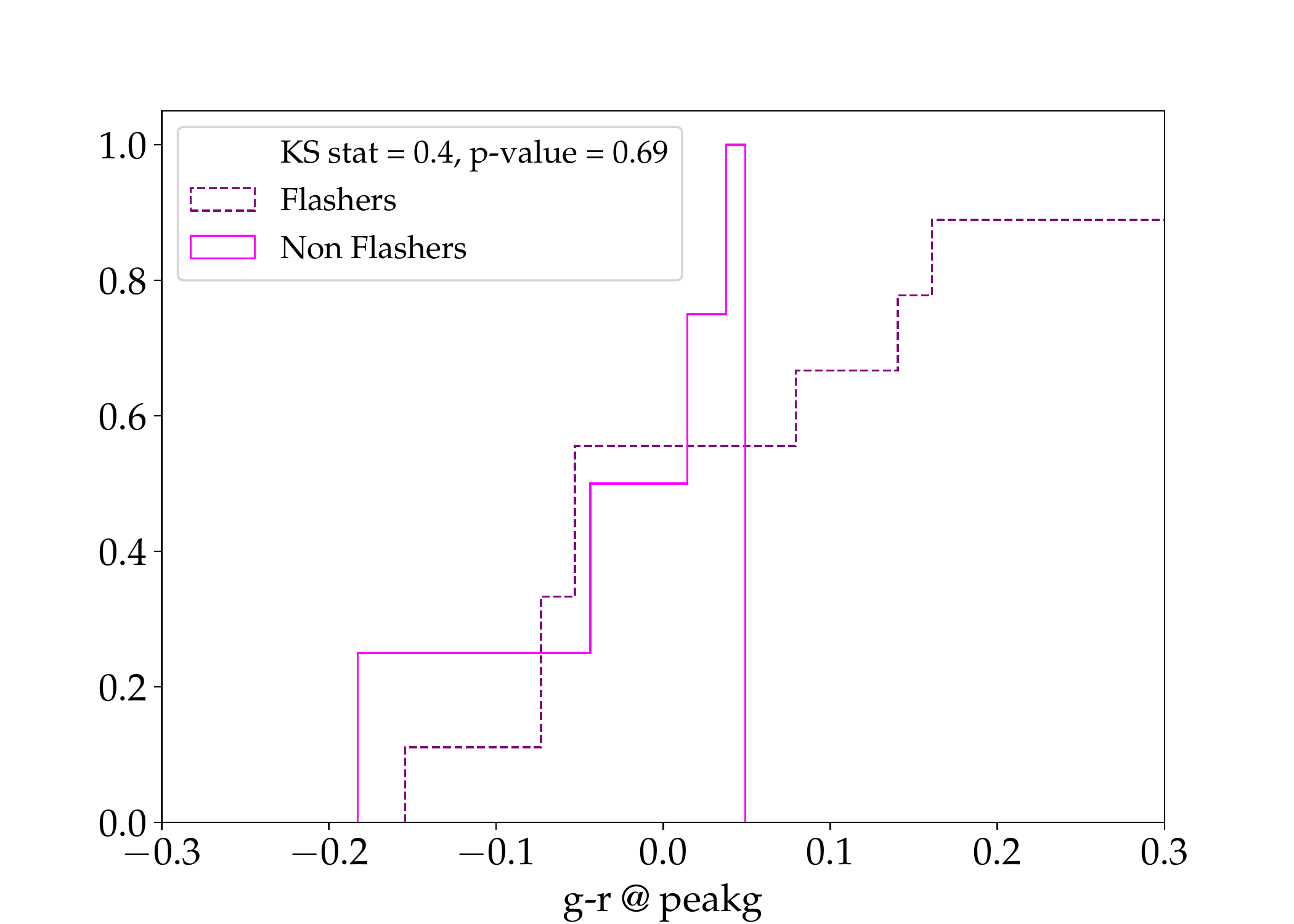}
    \caption{Top: color at peak of flasher (purple diamond) vs. non flasher (magenta triangle). Bottom: CDF of the two distributions, they are very likely to be drawn from the same distribution.}
    \label{fig:coloratpeak}
\end{figure}


\begin{table*}
\caption{Timescales of flash-ionisation features, peak absolute magnitudes in r band and g band. }
\hspace{-3cm}
\renewcommand{\arraystretch}{0.82}
\begin{tabular}{lllllllllllll}
\hline
IAU & Type & JD of & JD of  & $\tau$ & Error on & z & error& Band &  Peak & $\delta$M & Rise & error on \\
name &  & last flash & no flash  & &$\tau$ & z & on z &  &  Abs. Mag &  & time & rise time  \\
(SN) &  & [d] & [d]  & [d] & [d]&  &  &  &  [AB] & [AB] & [d] & [d]  \\
\hline
\hline
&  &  &  &  &  &  &  &  &  &  &  &  \\
2018grf & SN II & 2458379.5 & 2458386.5 & 5.39 & 3.50 & 0.054 & 0.0073 & r & -18.52 & 0.30 & 6.29 & 0.39 \\
&  &  &  &  &  &  &  & g & -18.58 & 0.30 & 4.88 & 0.29 \\
2019nvm & SN II & 2458715.5 & 2458717.5 & 1.88 & 1.00 & 0.018 & $<10^{-4}$ & r & -17.64 & 0.01 & 8.58 & 0.57 \\
&  &  &  &  &  &  &  & g & -17.47 & 0.02 & 7.68 & 1.22 \\
2018dfi & SN IIb & 2458307.5 & 2458311.5 & 2.25 & 2.05 & 0.031 & 0.0002& r  & -17.61 & 0.01 & 2.66 & 0.43 \\
&  &  &  &  &  &  &  & g & -17.76 & 0.02 & 1.72 & 0.44 \\
2020pni & SN II & 2459050.5 & 2459052.5 & 4.96 & 1.00 & 0.017 & $<10^{-4}$ & r & -18.25 & 0.01 & 11.01 & 0.27 \\
&  &  &  &  &  &  &  & g & -18.27 & $<10^{-2}$ & 6.27 & 0.53 \\
2020sic & SN II & 2459094.5 & 2459096.5 & 2.02 & 1.00 & 0.033 & 0.0001 & r & -17.87 & 0.29 & 12.37 & 1.57 \\
&  &  &  &  &  &  &  & g & - & - & - & - \\
2018dfc & SN II & 2458307.5 & 2458312.5 & 6.23 & 2.83 & 0.037 & 0.0001 & r  & -18.50 & 0.02 & 10.43 & 1.50 \\
&  &  &  &  &  &  &  & g  & -18.55 & 0.01 & 7.37 & 1.35 \\
2018fif & SN II & 2458351.5 & 2458353.5 & 1.62 & 1.00 & 0.017 & $<10^{-4}$ & r & -17.18 & 0.01 & 14.13 & 1.03 \\
&  &  &  &  &  &  &  & g & -17.02 & 0.03 & 10.04 & 1.69 \\
2018cyg & SN II & 2458295.5 & 2458296.5 & 1.28 & 0.50 & 0.012 & 0.0002 & r & -15.49 & 0.03 & 16.86 & 0.29 \\
&  &  &  &  &  &  &  & g & -14.51 & 0.03 & 10.37 & 0.30 \\
2020afdi & SN II & 2459071.5 & 2459072.5 & 1.30 & 0.50 & 0.024 & 0.0001 & r & -15.88 & 0.01 & 6.26 & 0.92 \\
&  &  &  &  &  &  &  & g & -15.94 & 0.01 & 4.72 & 0.55 \\
2019ust & SN II & 2458804.5 & 2458805.5 & 5.00 & 0.50 & 0.022 & $<10^{-4}$ & r & -18.17 & 0.02 & 12.47 & 1.65 \\
&  &  &  &  &  &  &  & g & -18.02 & $<10^{-2}$ & 15.15 & 0.11 \\
2020lfn & SN II & 2458998.5 & 2459001.5 & 4.18 & 1.50 & 0.044 & 0.0052 & r & -18.95 & 0.26 & 10.80 & 0.63 \\
&  &  &  &  &  &  &  & g & -18.96 & 0.26 & 9.13 & 0.29 \\
2018cug & SN II & 2458292.5 & 2458294.5 & 2.72 & 1.00 & 0.049 & 0.0024 & r & -18.20 & 0.11 & 10.46 & 0.48 \\
&  &  &  &  &  &  &  & g & -18.25 & 0.11 & 7.56 & 0.28 \\
2020ufx & SN II & 2459121.5 & 2459123.5 & 4.88 & 1.00 & 0.051 & 0.0021 & r & -18.93 & 0.09 & 10.84 & 0.70 \\
&  &  &  &  &  &  &  & g & -19.14 & 0.09 & 6.11 & 0.45 \\
2020pqv & SN II & 2459049.5 & 2459054.5 & 5.21 & 2.51 & 0.034 & $<10^{-4}$ & r & -18.02 & 0.01 & 24.65 & 0.87 \\
&  &  &  &  &  &  &  & g & -17.72 & $<10^{-2}$ & 4.56 & 0.16 \\
2018leh & SN II & 2458484.5 & 2458486.5 & 7.59 & 5.50 & 0.024 & $<10^{-4}$ & r & -18.01 & 0.01 & 14.70 & 0.15 \\
&  &  &  &  &  &  &  & g & -18.05 & $<10^{-2}$ & 12.49 & 0.14 \\
2020wol & SN II & 2459143.5 & 2459156.5 & 13.56 & 6.51 & 0.050 & 0.0100 & r  & -18.92 & 0.43 & 16.25 & 0.77 \\
&  &  &  &  &  &  &  & g & -19.05 & 0.43 & 10.97 & 0.64 \\
2019qch & SN II & 2458750.5 & 2458751.5 & 14.64 & 1.13 & 0.024 & 0.0014 & r  & -18.23 & 0.13 & 19.74 & 1.03 \\
&  &  &  &  &  &  &  & g & -18.40 & 0.13 & 15.67 & 1.04 \\
2019mor & SN II & 2458699.5 & 2458703.5 & 8.28 & 2.50 & 0.019 & 0.0001 & r & -17.18 & 0.02 & 12.97 & 1.53 \\
&  &  &  &  &  &  &  & g & -17.35 & 0.01 & 11.49 & 1.50 \\
2019lkw & SN II & 2458690.5 & 2458696.5 & 17.15 & 3.10 & 0.073 & 0.0021& r  & -20.13 & 0.06 & 14.03 & 0.83 \\
&  &  &  &  &  &  &  & g & -20.37 & 0.06 & 13.37 & 0.81 \\
&  &  &  &  &  &  &  &  &  &  &  &  \\
\hline
\end{tabular}
\renewcommand{\arraystretch}{1}
\label{tab:timnescaleflash}

\end{table*}

\subsection{Duration of flash features}
We estimate the flash timescale as the time from the estimated explosion date until the half-time between the last spectrum still showing a He II (4686Å) line and the first spectrum not showing He II line anymore, see Table \ref{tab:timnescaleflash}.  We designate these two spectra as bounding spectra. We look for correlations of flash-feature timescales against peak absolute magnitude and rise time, considering all the infant candidates which showed flash-features and from which we could estimate a timescale (15 out of 29). We disqualify those for which the time between the last spectrum showing the He II emission line and the first with no line or broad feature was longer than 14 days. We also disqualify candidates whose first spectrum with no He II line had a SNR lower than 15. We designate this subsample as the golden flasher sample, see Figure \ref{fig:subsamps} in the Appendix.\\ 

\begin{figure}[!h]
\hspace{-0.5cm}
    \includegraphics[trim=10 30 30 50,clip,width = 1.1\columnwidth]{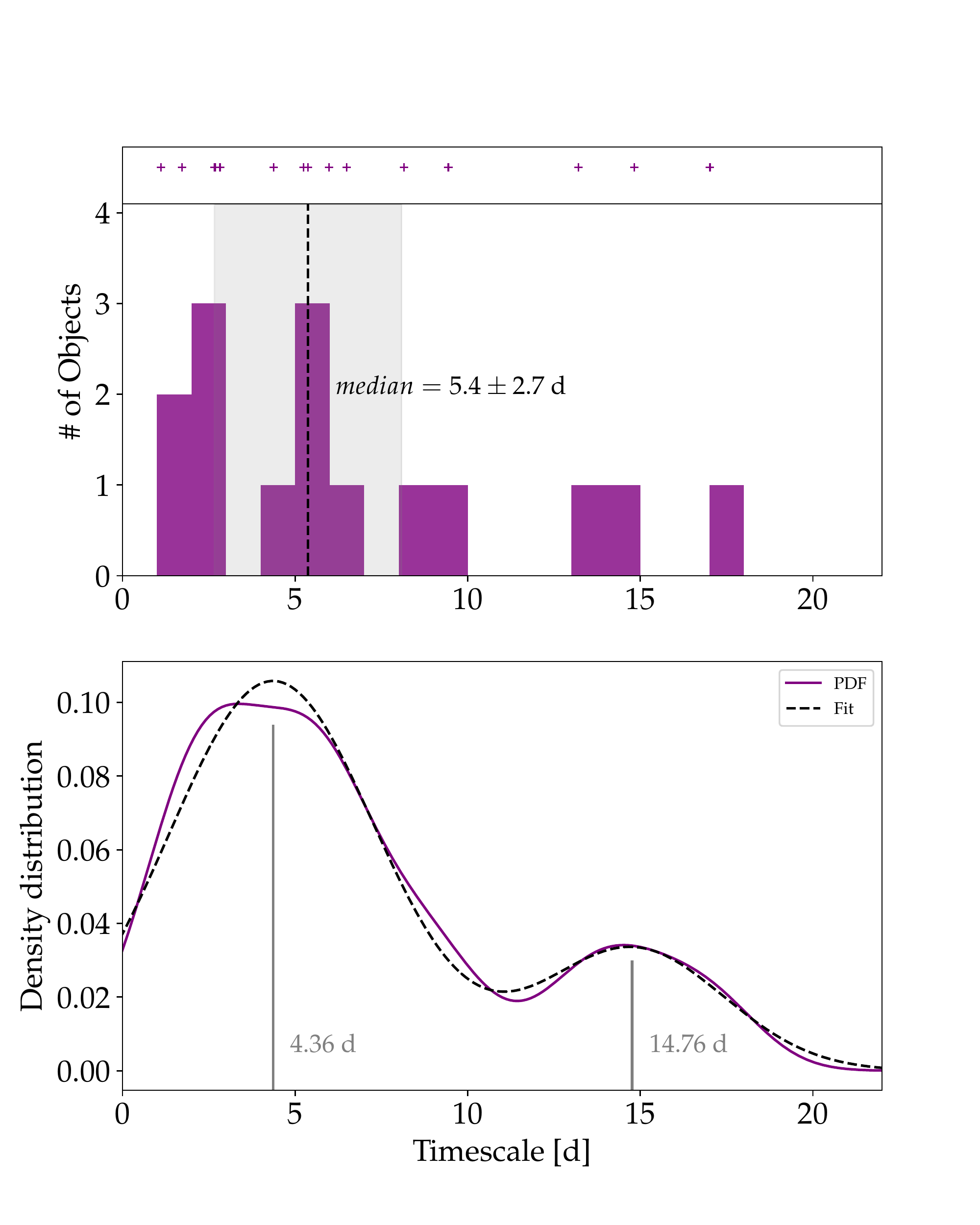}
    \caption{Histogram of the timescales for flash ionisation features over a sample of 15 candidates. }
    \label{fig:timescaledistribution}
\end{figure}

The distribution of flash timescales (Fig. \ref{fig:timescaledistribution}) shows that most SNe have a characteristic timescale of flash features shorter than $5.4\pm2.7$ days. We extract the probability distribution function of the timescale of flash features, using Kernel Density estimation. We represent each measurement as a Gaussian kernel function, whose mean is the value of the measurement. We chose a fixed-width kernel. The width is the median value of all the estimated errors, i.e. $\sigma = 1.48 $d. The resulting estimated density is the bottom panel in Figure \ref{fig:timescaledistribution}. We fit two Gaussians and find two clusters at 4.36 d and 14.76 d respectively. The size of our sample is however too small to interpret this probability density function (PDF) meaningfully. While a single Gaussian cannot reproduce the timescale measurements at $\geq 12 $ days, it is not yet possible to tell whether the longer-lived flasher belong to a distinct family (which would show different physical properties in interaction with the CSM); or if the PDF is a skewed normal distribution, hence making longer-lived flasher a rarer population.

\begin{figure}[h!]
\hspace{-1cm}
    \centering
    \includegraphics[width = 1.1\columnwidth]{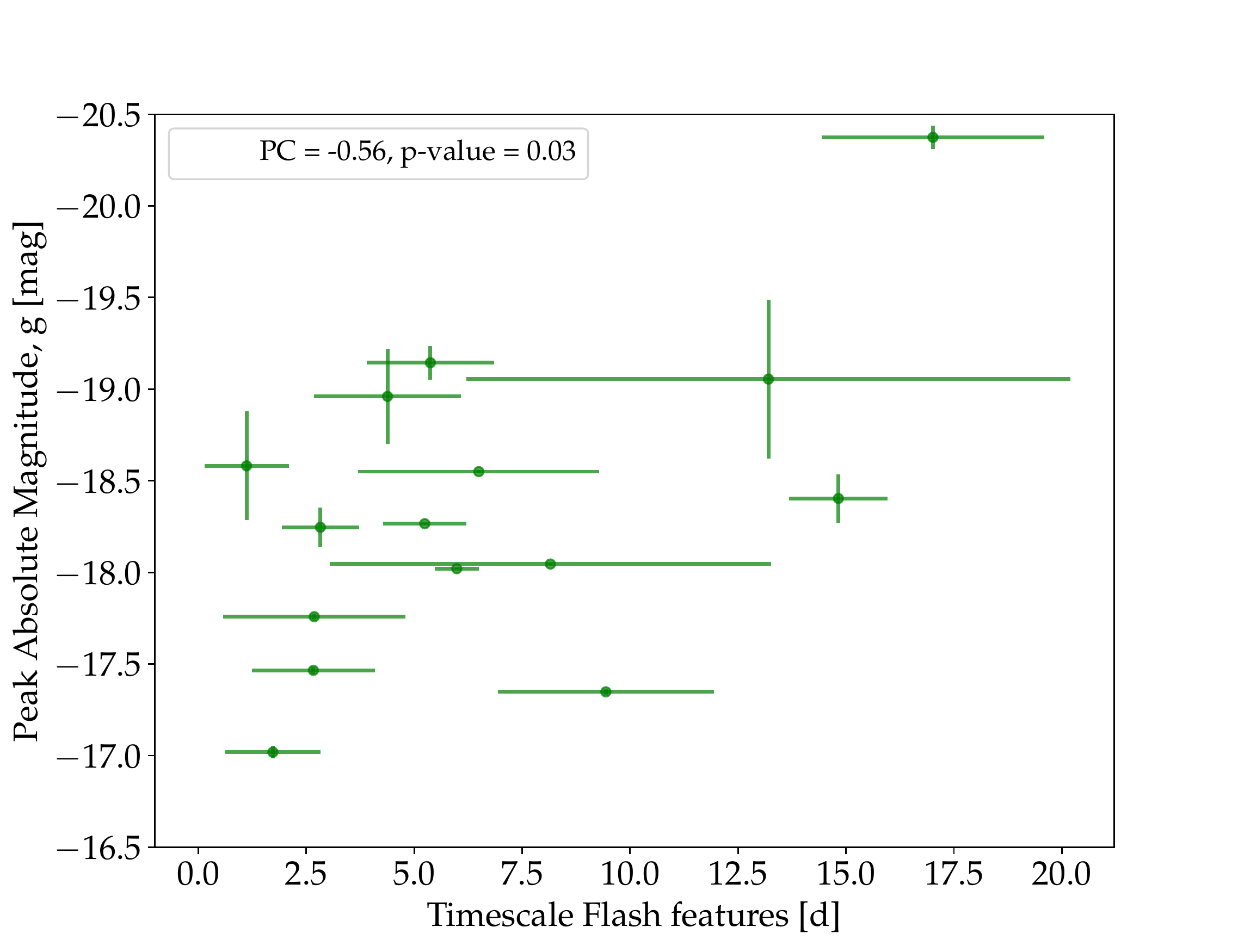}\\
\hspace{-1cm}
    \includegraphics[width = 1.1\columnwidth]{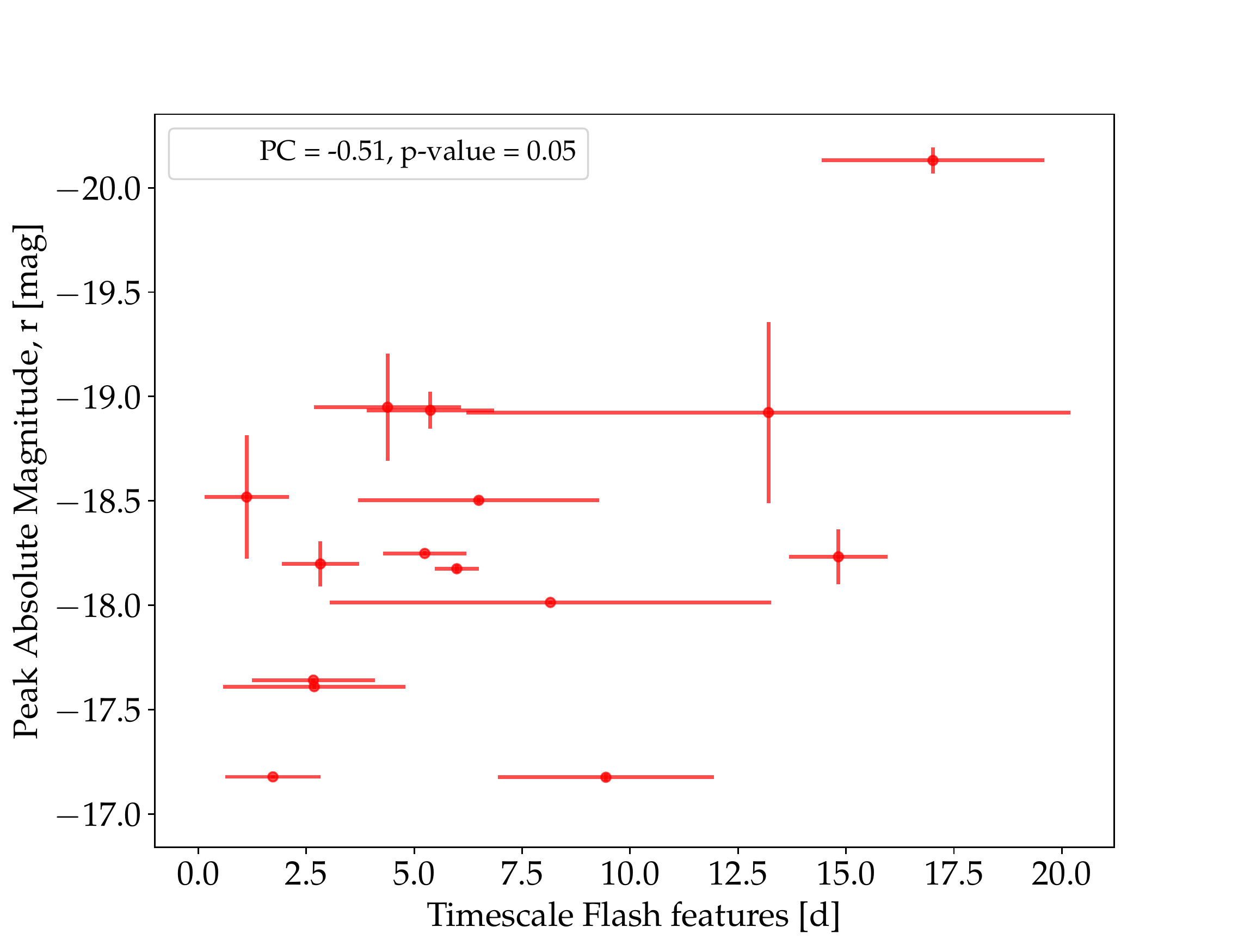}
    \caption{Flash timescale vs. peak magnitude }
    \label{fig:timescalepeak}
\end{figure}

\begin{figure}[h!]
\hspace{-1cm}
    \centering
     \includegraphics[width = 1.1\columnwidth]{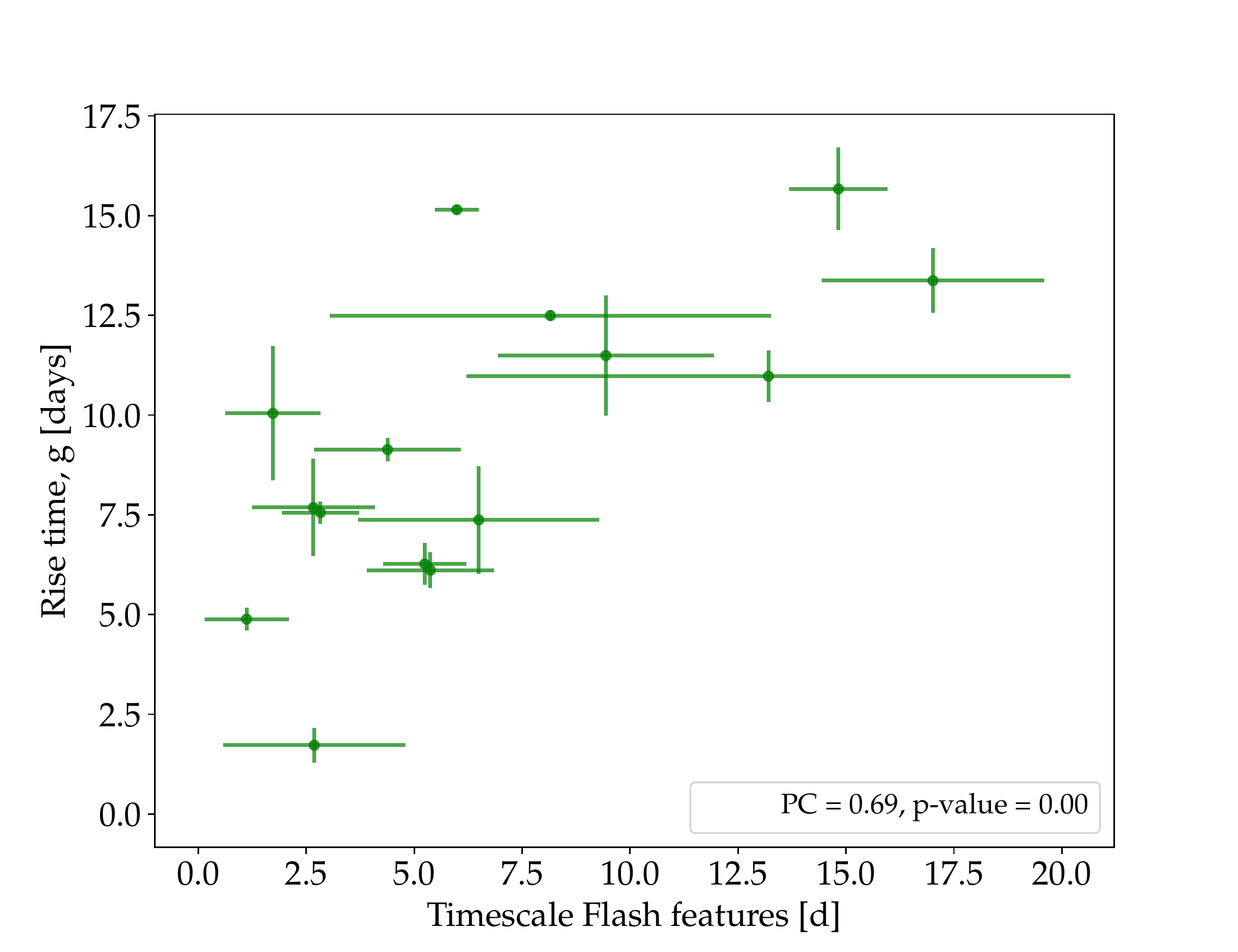}\\
\hspace{-1cm}
    \includegraphics[width = 1.1\columnwidth]{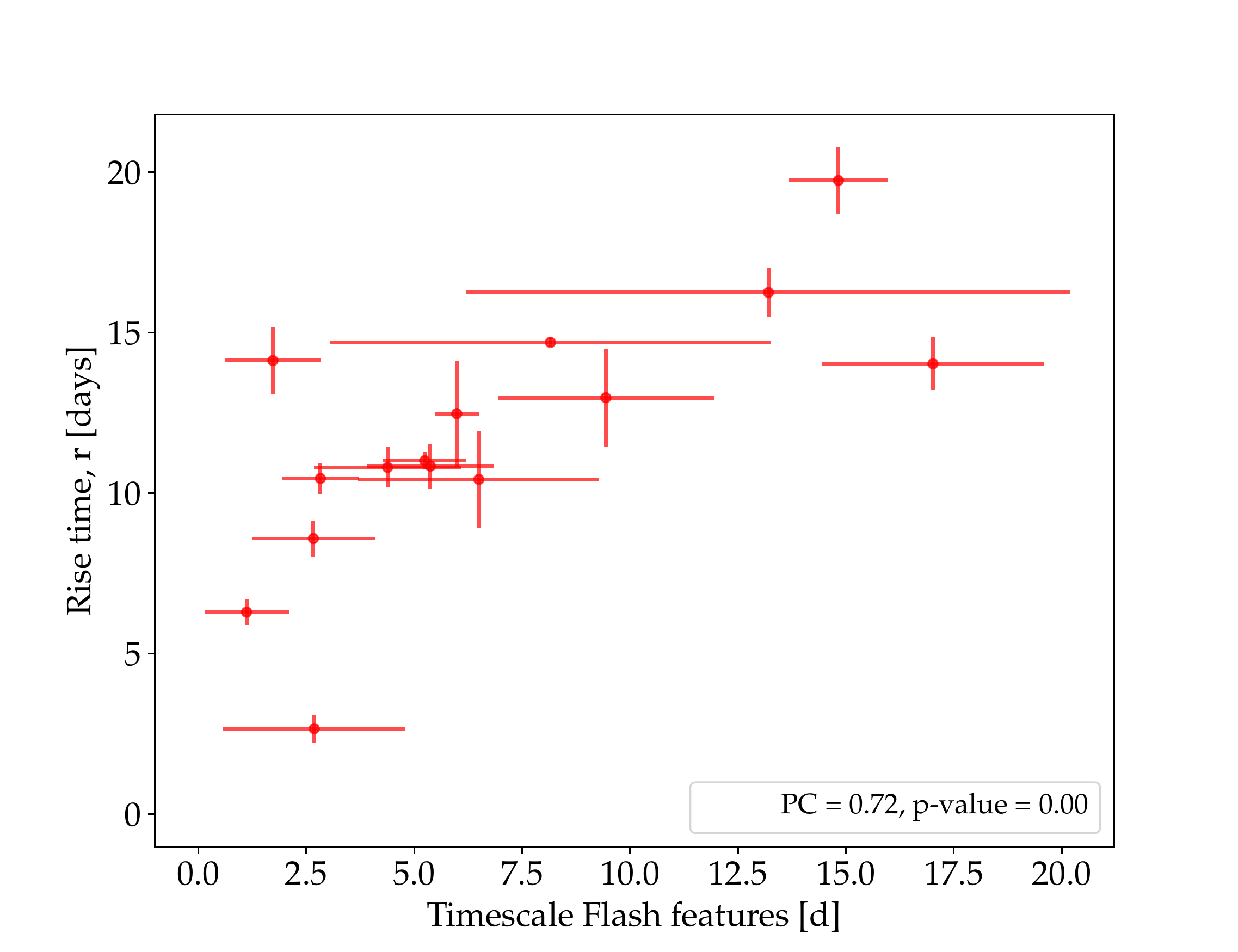}
    \caption{Flash timescale vs. rise time }
    \label{fig:timescalerise}
\end{figure}

We find a partial correlation between the flash features time scale and the peak absolute magnitude: the Pearson correlation coefficient is $-0.51$ in $r$ band and $-0.56$ in $g$ band, see Fig \ref{fig:timescalepeak}. The probability for an uncorrelated distribution to generate such a coefficient is low, with a p-value of $\lesssim 0.05$.  However, it seems that the long-lived flasher population, with timescales longer than 10 days, are driving this correlation. The low number of observations in our sample does not allow us to make any strong conclusion. 
The flash feature timescale is correlated with the rise time, with a Pearson coefficient of 0.69 in the $g$ band and 0.72 in the $r$ band. The p-values in both cases are $\lesssim0.05$, hence excluding the chance that an uncorrelated distribution could generate such coefficient. We observe that the longer the flash timescale is, the longer the rise time is to a brighter peak. \\

Our work suggests, for the first time, that there exist a characteristic timescale which may separate events where the CSM does not significantly influence the brightness of the light curve ($<9$ days, common flashers), and events where the CSM does ($>9$ days, long-lived flash features). In the latter case, such events display brighter peaks and a slower rise to the peak. This population of long-lasting flashers could be considered as the low-CSM mass tail of the population of SNe IIn.


\section{Conclusion}\label{conclusion}
We report the results of the search for flash-ionisation features in infant, hydrogen-rich SNe during the first phase of ZTF (from March 2018 to December 2020). We collected 148 such objects (at a rate of once to twice a week) and obtained rapid follow-up spectroscopy within $2$\,d from estimated explosion for 25 of the SNe classified as spectroscopically-normal SNe II. \\
Fast response spectroscopic facilities were essential, such as the SEDm with which we obtained 13 of those spectra. However, 7 were disqualified due to their low S/N ratio. We corroborated our previous results, see \cite{bruch2021}, that flash-ionisation features occur in at least 30\%, and likely most, hydrogen-rich SNe. This implies that a confined CSM is common around hydrogen-rich SN progenitors. \\
We also investigated the early light-curve behaviour (i.e. rise times, peak magnitudes and color at peak) of 17 events which have a spectrum less than two days from estimated explosion. We find that there is no significant difference between candidates showing flash-ionisation features and those who do not at early time. We hence conclude that the confined CSM of most SNe is not massive enough to contribute extra energy to the light curve at early time. \\
We present for the first time a sample of flash ionisation events with sequences measuring the timescale of disappearance of flash-ionisation features. Typical flash events last for $\approx 5$ days. A rarer population of flashers have timescales above $10$ days from the estimated explosion time. This population have longer rise times and reach brighter peak magnitudes. We hypothesize that this group bridges between spectroscopically-normal SNe II and strongly interacting SNe IIn. It is not clear yet if the distribution of timescale of flash-ionisation features is bimodal (i.e. two distinct population of typical flashers and long-lasting flashers) or a skewed normal distribution (i.e. long lasting flashers are just rarer). Our results also question the CSM properties as well as the regime of CSM interaction in which flash feature are. \\
Since no significant rise time or peak absolute magnitude difference is found between events with flash features and those without, we could hypothesize that no significant energy conversion is taking place between the ejecta and the CSM. Thus making the regime of interaction for common flash ionisation shockless. Events for which higher peak magnitudes and longer rise times are recorded could thus also represent a transitional population from flash ionisation to shock ionisation. These different regimes of interaction would be a direct consequence of the CSM properties, i.e. from optically thin and confined for regular flash ionisation to optically thick and extended for shock ionisation. \\
Our results motivate the systematic acquisition of series of spectra for young, hydrogen-rich SNe. While we have established that the presence of CSM is common around massive star progenitors, the properties of the CSM, such as density profile, composition, compactness were not studied over big samples. Mapping these properties will shed light on the environment and circumstances prior to the explosion of hydrogen-rich SN progenitors. Such studies are currently ongoing in partnership with the ePessto$+$ survey, using the EFOSC2 (R$\approx 390$) on the NTT spectrographs.  They should be continued  with the upcoming instrument SoXs, a mid-resolution spectrograph (R$\approx 4500 $) with high throughput \cite{rubin2020}. This instrument will be dedicated to transient science with Target of Opportunity observing strategy, hence allowing for more systematic rapid-response spectroscopic follow up. The commissioning of SoXs is planned to start in September 2023.


\newpage

\begin{acknowledgments}
AGY's research is supported by the EU via ERC grant No. 725161, the ISF GW excellence center, an IMOS space infrastructure grant and BSF/Transformative and GIF grants, as well as The Benoziyo Endowment Fund for the Advancement of Science, the Deloro Institute for Advanced Research in Space and Optics, The Veronika A. Rabl Physics Discretionary Fund, Paul and Tina Gardner, Yeda-Sela and the WIS-CIT joint research grant;  AGY is the recipient of the Helen and Martin Kimmel Award for Innovative Investigation.
NLS is funded by the Deutsche Forschungsgemeinschaft (DFG, German Research Foundation) via the Walter Benjamin program – 461903330.
SED Machine is based upon work supported by the National Science Foundation under Grant No. 1106171.%
The ztfquery code was funded by the European Research Council (ERC) under the European Union's Horizon 2020 research and innovation programme (grant agreement No. 759194 - USNAC, PI: Rigault).%

The ZTF forced-photometry service was funded under the Heising-Simons Foundation grant \#12540303 (PI: Graham).%

Based on observations obtained with the Samuel Oschin 48-inch Telescope at the Palomar Observatory as part of the Zwicky Transient Facility project. ZTF is supported by the National Science Foundation under Grant No. AST-1440341 and a collaboration including Caltech, IPAC, the Weizmann Institute for Science, the Oskar Klein Center at Stockholm University, the University of Maryland, the University of Washington, Deutsches Elektronen-Synchrotron and Humboldt University, Los Alamos National Laboratories, the TANGO Consortium of Taiwan, the University of Wisconsin at Milwaukee, and Lawrence Berkeley National Laboratories. Operations are conducted by COO, IPAC, and UW. %

Based on observations made with the Nordic Optical Telescope, owned in collaboration by the University of Turku and Aarhus University, and operated jointly by Aarhus University, the University of Turku and the University of Oslo, representing Denmark, Finland and Norway, the University of Iceland and Stockholm University at the Observatorio del Roque de los Muchachos, La Palma, Spain, of the Instituto de Astrofisica de Canarias.%

Based on observations obtained at the international Gemini Observatory, a program of NSF’s NOIRLab, which is managed by the Association of Universities for Research in Astronomy (AURA) under a cooperative agreement with the National Science Foundation. on behalf of the Gemini Observatory partnership: the National Science Foundation (United States), National Research Council (Canada), Agencia Nacional de Investigaci\'{o}n y Desarrollo (Chile), Ministerio de Ciencia, Tecnolog\'{i}a e Innovaci\'{o}n (Argentina), Minist\'{e}rio da Ci\^{e}ncia, Tecnologia, Inova\c{c}\~{o}es e Comunica\c{c}\~{o}es (Brazil), and Korea Astronomy and Space Science Institute (Republic of Korea). %

This research has made use of the NASA/IPAC Extragalactic Database (NED), which is funded by the National Aeronautics and Space Administration and operated by the California Institute of Technology.

\end{acknowledgments}

\bibliography{infant_sne_bibliography}

\appendix






\section{The hydrogen rich sample}

\begin{table*}[h!]
\caption{Hydrogen rich normal SNe II (part 1/3)}
\hspace{-2.5cm}
\begin{tabular}{lllllllllll}
\hline
IAU & ZTF & Type & Explosion & Error & Last & First & RA & DEC & First & Flasher \\
name & name &  & JD date &  & Non  & detection & (median) & (median) & spectrum &  \\
& (ZTF) &  &  &  & detection &  &  &  &  &  \\
& &  & [d] & [d] & [d] & [d] & [degrees] & [degrees] & [d] &  \\
\hline
\hline
2018iuq  & 18acqwdla & SN II & 2458443.832 & 0.042 & -0.043 & 0.042 & 106.472662 & 12.8929375 & 0.105 & no \\
2018grf  & 18abwlsoi & SN II & 2458377.609 & 0.003 & -0.869 & 0.021 & 261.897614 & 71.530251 & 0.142 & yes \\
2020acbm & 20acwgxhk & SN II & 2459193.562 & 0.030 & -0.850 & 0.125 & 40.0741593 & 2.4270671 & 0.167 & no \\
2019nvm  & 19abqhobb & SN II & 2458714.625 & 0.006 & -0.883 & 0.038 & 261.4111 & 59.4467303 & 0.167 & yes \\
2020qvw  & 20abqkaoc & SN II & 2459067.290 & 0.490 & -0.490 & 0.490 & 250.983335 & 77.879897 & 0.710 & no \\
2020pni  & 20ablygyy & SN II & 2459046.539 & 0.031 & -0.785 & 0.159 & 225.958184 & 42.1140315 & 0.864 & yes \\
2020sic  & 20abxyjtx & SN II & 2459093.484 & 0.008 & -1.768 & 0.150 & 236.937978 & 28.6403193 & 0.891 & yes \\
2018cxn  & 18abckutn & SN II & 2458289.758 & 0.015 & 0.000 & 0.107 & 237.026897 & 55.7148553 & 0.990 & no \\
2018dfc  & 18abeajml & SN II & 2458303.773 & 1.332 & -0.976 & 0.026 & 252.03236 & 24.3040949 & 1.021 & yes \\
2019omp  & 19abrlvij & SN II & 2458718.809 & 0.010 & 0.001 & 0.841 & 260.142987 & 51.6327799 & 1.054 & no \\
2019ewb  & 19aatqzrb & SN II & 2458606.787 & 0.003 & -0.899 & 0.012 & 221.652383 & 56.2342197 & 1.083 & no \\
2020dyu  & 20aasfhia & SN II & 2458912.697 & 0.012 & -0.727 & 0.083 & 184.913045 & 33.0403926 & 1.121 & no \\
2018fif  & 18abokyfk & SN II & 2458350.877 & 0.008 & -0.976 & 0.013 & 2.360629 & 47.3540827 & 1.129 & yes \\
2020abbo & 20acuaqlf & SN II & 2459181.374 & 0.060 & -1.703 & 0.234 & 357.775211 & 6.9424927 & 1.206 & no \\
2020mst  & 20abfcdkj & SN II & 2459013.701 & 0.008 & -0.881 & 0.049 & 281.793965 & 60.4968018 & 1.299 & no \\
2020dya  & 20aasijew & SN II & 2458912.504 & 0.454 & -0.454 & 0.454 & 216.905399 & 69.6864096 & 1.379 & no \\
2020sjv  & 20abybeex & SN II & 2459094.200 & 0.498 & -0.499 & 0.498 & 260.769541 & 55.0724721 & 1.508 & no \\
2018cyg  & 18abdbysy & SN II & 2458294.724 & 0.002 & 0.057 & 0.981 & 233.535367 & 56.6968577 & 1.676 & yes? \\
2020afdi & 20abqwkxs & SN II & 2459070.698 & 0.029 & -0.899 & 0.006 & 224.868111 & 73.8986784 & 1.693 & yes \\
2020uim  & 20acfdmex & SN II & 2459118.846 & 0.005 & -0.985 & 0.002 & 28.1887405 & 36.6231594 & 1.719 & no \\
2018egh  & 18abgqvwv & SN II & 2458312.710 & 0.001 & 0.128 & 1.020 & 254.316401 & 31.9631992 & 1.859 & yes? \\
2020xhs  & 20acknpig & SN II & 2459139.073 & 0.059 & -0.206 & 1.685 & 30.7428678 & 45.0202856 & 1.888 & no \\
2019ikb  & 19abbwfgp & SN II & 2458661.817 & 3.554 & -0.974 & 0.003 & 258.323795 & 43.7843194 & 1.942 & no \\
2019ust  & 19acryurj & SN II & 2458799.997 & 0.032 & -0.192 & 0.793 & 13.5933959 & 31.6701819 & 1.994 & yes \\
\hline
2020lfn  & 20abccixp & SN II & 2458995.816 & 0.002 & 0.004 & 0.954 & 246.737034 & 20.2459056 & 2.011 & yes \\
2019dky  & 19aapygmq & SN II & 2458584.778 & 4.036 & 0.093 & 0.978 & 210.421485 & 38.5103291 & 2.047 & no \\
2019odf  & 19abqrhvy & SN II & 2458714.844 & 0.008 & 0.028 & 1.073 & 342.186213 & 27.5718269 & 2.139 & no \\
2019gmh  & 19aawgxdn & SN II & 2458633.768 & 5.618 & 0.078 & 1.940 & 247.763189 & 41.1539613 & 2.166 & yes? \\
2018bqs  & 18aarpttw & SN II & 2458246.812 & 0.001 & -1.963 & 0.010 & 247.259916 & 43.6268251 & 2.188 & no \\
2020uhf  & 20aceyolc & SN II & 2459118.795 & 0.004 & -0.903 & 0.001 & 44.102817 & 38.1871607 & 2.205 & yes \\
2018cug  & 18abcptmt & SN II & 2458290.779 & 0.022 & -0.038 & 0.085 & 267.329908 & 49.412409 & 2.221 & yes \\
2019oxn  & 19abueupg & SN II & 2458724.584 & 0.012 & -0.774 & 0.066 & 267.80329 & 51.3825496 & 2.281 & no \\
2019szo  & 19acgbkzr & SN II & 2458775.334 & 0.012 & -0.619 & 0.471 & 4.9860264 & 15.0933857 & 2.282 & no \\
2020ufx  & 20acedqis & SN II & 2459117.623 & 0.010 & -0.784 & 0.020 & 322.652706 & 24.6737523 & 2.377 & yes \\
2020umi  & 20acfkzcg & SN II & 2459119.496 & 0.477 & -0.478 & 0.477 & 115.76978 & 50.2887543 & 2.413 & no \\
2019dlo  & 19aapvltt & SN II & 2458583.799 & 0.101 & -0.810 & 0.087 & 267.6325156 & 58.6245046 & 2.701 & no \\
2018fsm  & 18absldfl & SN II & 2458362.965 & 0.009 & 0.005 & 0.915 & 33.5997569 & 30.811935 & 2.825 & no \\
\hline
\end{tabular}
\label{tab:allSNII}
\end{table*}

\begin{table*}[h!]
\caption{Hydrogen rich normal SNe II (part 2/3)}
\hspace{-2.5cm}
\begin{tabular}{lllllllllll}
\hline
IAU & ZTF & Type & Explosion & Error & Last & First & RA & DEC & First & Flasher \\
name & name &  & JD date &  & Non  & detection & (median) & (median) & spectrum &  \\
& (ZTF) &  &  &  & detection &  &  &  &  &  \\
& &  & [d] & [d] & [d] & [d] & [degrees] & [degrees] & [d] &  \\
\hline
\hline
2020zpt  & 20acqexmr & SN II & 2459166.900 & 0.025 & -1.966 & 0.014 & 57.9034378 & 43.6980162 & 2.828 & no \\
2020xva  & 20aclvtnk & SN II & 2459141.653 & 2.184 & 1.965 & 1.989 & 263.035128 & 53.6539888 & 2.847 & no \\
2018gts  & 18abvvmdf & SN II & 2458373.738 & 0.003 & 0.000 & 0.896 & 249.197462 & 55.7357948 & 2.881 & yes \\
2018bge  & 18aaqkoyr & SN II & 2458242.776 & 0.198 & -0.126 & 0.909 & 166.066683 & 50.0306395 & 2.908 & no \\
2020uao  & 20accrldu & SN II & 2459114.783 & 0.004 & -0.893 & 0.041 & 17.1983968 & 27.0450181 & 2.912 & no \\
2020yui  & 18aadsuxd & SN II & 2459154.018 & 0.007 & 0.000 & 1.947 & 129.533971 & 31.667916 & 2.921 & no \\
2019eoh  & 19aatqzim & SN II & 2458604.976 & 0.422 & -3.194 & 1.715 & 195.955635 & 38.2891552 & 2.930 & no \\
2020ifv  & 20aawgrcu & SN II & 2458963.957 & 0.024 & -0.007 & 0.943 & 310.650913 & 76.7817657 & 2.939 & no \\
2020pqv  & 20abmoakx & SN II & 2459046.791 & 0.164 & -0.081 & 1.968 & 220.49818 & 8.46272355 & 2.989 & yes \\
2018leh  & 18adbmrug & SN II & 2458482.405 & 0.039 & -1.603 & 0.294 & 61.2637726 & 25.2619268 & 3.044 & yes \\
2020uqx  & 20acgided & SN II & 2459123.544 & 0.049 & -0.800 & 0.153 & 326.826979 & 32.0957996 & 3.143 & yes \\
2020wol  & 20acjbhhp & SN II & 2459136.445 & 0.291 & -1.615 & 0.385 & 29.8613506 & 30.726751 & 3.240 & yes \\
2020dbg  & 20aapycrh & SN II & 2458900.752 & 0.056 & -0.884 & 0.007 & 164.245241 & 43.0768461 & 3.248 & no \\
2019twk  & 19aclobbu & SN II & 2458788.242 & 0.087 & -1.482 & 0.528 & 35.7720108 & 46.8824189 & 3.407 & no \\
2020wog  & 20aciwrpn & SN II & 2459135.971 & 0.067 & -1.251 & 0.689 & 328.182556 & 33.6561827 & 3.797 & no \\
2020ykb  & 20acocohy & SN II & 2459149.954 & 0.042 & -0.012 & 0.943 & 64.0382399 & -25.474303 & 3.865 & no \\
2020iez  & 20aavvaup & SN II & 2458962.615 & 0.007 & -7.898 & 0.065 & 147.118273 & 50.9224955 & 3.885 & no \\
2019ssi  & 19acftfav & SN II & 2458773.704 & 0.053 & 0.048 & 0.993 & 352.733873 & 15.4916278 & 3.890 & no \\
2020sje  & 20abxmwwd & SN II & 2459089.911 & 0.041 & 0.049 & 0.959 & 19.3414942 & 44.1948692 & 3.963 & no \\
2020dbd  & 20aapjiwl & SN II & 2458899.750 & 0.027 & -0.940 & 0.011 & 142.988659 & 33.2096657 & 4.125 & no \\
2019lnl  & 19abgrmfu & SN II & 2458681.566 & 0.020 & -0.811 & 0.117 & 255.526652 & 32.9977641 & 4.154 & no \\
2020iyi  & 20aaxunbm & SN II & 2458969.452 & 0.045 & -0.712 & 0.231 & 152.764995 & 54.369596 & 4.233 & no \\
2018iua  & 18acploez & SN II & 2458439.478 & 0.469 & -0.469 & 0.468 & 130.037293 & 68.9031911 & 4.303 & no \\
2020ovk  & 20ablklei & SN II & 2459042.418 & 0.024 & -1.433 & 0.443 & 358.573861 & 26.3267084 & 4.472 & no \\
2020cvy  & 20aaophpu & SN II & 2458895.697 & 0.030 & 1.026 & 2.993 & 120.205292 & 27.4985715 & 4.943 & no \\
2020ult  & 20acfkyll & SN II & 2459119.496 & 0.477 & -0.478 & 0.477 & 106.885226 & 48.9002151 & 5.160 & no \\
2020rjd  & 20absgwch & SN II & 2459075.643 & 0.033 & -1.735 & 0.265 & 359.74376 & 3.7404556 & 5.214 & no? \\
2020yts  & 20acongti & SN II & 2459153.623 & 0.002 & 0.116 & 1.068 & 338.614616 & 25.0352627 & 5.223 & no \\
2019tjt  & 19acignlo & SN II & 2458782.626 & 2.238 & -0.966 & 0.004 & 4.2653086 & 31.5725235 & 5.374 & yes \\
2020rth  & 20abupxie & SN II & 2459080.475 & 0.485 & -0.485 & 0.485 & 52.1135451 & -5.2545457 & 5.525 & no \\
2018fpb  & 18abqyvzy & SN II & 2458357.826 & 0.005 & -0.856 & 0.089 & 359.9284 & 34.3444464 & 5.873 & no \\
2020dbn  & 20aaqbach & SN II & 2458899.926 & 0.009 & 0.068 & 0.911 & 187.036133 & 20.178184 & 5.990 & no \\
2018gvn  & 18abyvenk & SN II & 2458385.618 & 0.509 & -0.858 & 0.002 & 273.976407 & 44.6964598 & 6.114 & no \\
2019oba  & 19abpyqog & SN II & 2458711.672 & 0.003 & -0.824 & 1.079 & 299.264485 & 50.1889432 & 6.129 & no \\
2020pnn  & 20abmihnc & SN II & 2459044.747 & 0.026 & -2.955 & 0.000 & 271.019773 & 22.0370976 & 6.193 & no \\
2019qch  & 19abyuzch & SN II & 2458736.360 & 1.017 & 2.281 & 3.280 & 277.30754 & 41.0423427 & 6.265 & yes \\
2019cem  & 19aamtwiz & SN II & 2458559.435 & 0.083 & -0.581 & 0.334 & 199.151274 & 35.516058 & 6.345 & no \\
2018clq  & 18aatlfus & SN II & 2458248.900 & 0.958 & -0.959 & 0.958 & 257.176414 & 28.5206041 & 6.925 & no \\
2019mor  & 19abjsmmv & SN II & 2458693.223 & 1.500 & -1.500 & 1.500 & 234.658542 & 36.9586362 & 7.439 & yes \\
2019vdl  & 19actnwtn & SN II & 2458804.018 & 0.972 & -0.972 & 0.971 & 142.38251 & 44.4222435 & 7.857 & no \\
2019oot  & 19abrbmvt & SN II & 2458716.891 & 0.004 & -0.004 & 0.004 & 345.069535 & 24.7855163 & 8.014 & no \\
2020jmb  & 20aayrobw & SN II & 2458977.190 & 0.470 & -0.470 & 0.470 & 142.804994 & 38.2540186 & 8.810 & no \\
\hline
\end{tabular}
\end{table*}

\begin{table*}[h!]
\caption{Hydrogen rich normal SNe II (part 3/3)}
\hspace{-2.5cm}
\begin{tabular}{lllllllllll}
\hline
IAU & ZTF & Type & Explosion & Error & Last & First & RA & DEC & First & Flasher \\
name & name &  & JD date &  & Non  & detection & (median) & (median) & spectrum &  \\
& (ZTF) &  &  &  & detection &  &  &  &  &  \\
& &  & [d] & [d] & [d] & [d] & [degrees] & [degrees] & [d] &  \\
\hline
\hline
2019tbq  & 19acgzzea & SN II & 2458776.871 & 0.152 & -0.939 & 0.062 & 77.9474245 & 52.5389439 & 8.845 & no \\
2018inm  & 18achtnvk & SN II & 2458432.878 & 2.559 & 0.073 & 1.983 & 96.1686908 & 46.5038794 & 9.047 & no \\
2018ccp  & 18aawyjjq & SN II & 2458262.857 & 0.024 & -0.910 & 0.049 & 263.058847 & 36.0739975 & 9.143 & no \\
2019rsw  & 19accbeju & SN II & 2458757.761 & 0.037 & -0.847 & 0.174 & 37.8938775 & 24.8167672 & 9.164 & no \\
2020iho  & 20aawbzlo & SN II & 2458964.616 & 0.020 & -0.862 & 0.082 & 166.411236 & 30.8325541 & 9.191 & no \\
2019lkw  & 19abgpgyp & SN II & 2458676.348 & 0.790 & 0.475 & 1.420 & 256.287926 & 33.4425697 & 9.544 & yes \\
2020jww  & 20aazpphd & SN II & 2458982.874 & 1.077 & -1.077 & 1.076 & 242.714935 & 27.1616704 & 9.988 & no \\
2018lth  & 18aayxxew & SN II & 2458276.725 & 0.007 & 0.015 & 1.975 & 197.139654 & 45.9862178 & 10.091 & no \\
2020buc  & 18aaaibml & SN II & 2458881.397 & 0.475 & -0.476 & 0.475 & 152.130328 & 9.2397339 & 10.118 & yes \\
2019mge  & 19abjioie & SN II & 2458691.398 & 0.213 & 0.410 & 1.310 & 259.203094 & 39.1480677 & 11.405 & yes \\
2018iwe  & 18abufaej & SN II & 2458368.807 & 0.002 & -0.906 & 0.001 & 4.4825224 & 12.091568 & 12.065 & no \\
2020lam  & 20abbpkpa & SN II & 2458992.833 & 2.776 & 0.000 & 2.082 & 254.098077 & 26.8138571 & 12.112 & no \\
2019mkr  & 19abjrjdw & SN II & 2458694.510 & 0.044 & -1.792 & 0.167 & 257.774102 & 5.8520255 & 12.200 & no \\
2020rhg  & 20abqferm & SN II & 2459065.415 & 0.505 & -0.505 & 0.505 & 9.7067662 & 3.4036991 & 12.494 & no \\
2018iug  & 18acnmifq & SN II & 2458437.855 & 0.040 & -1.022 & 0.013 & 101.979329 & 67.9163152 & 12.834 & no \\
2020smm  & 20abykfsr & SN II & 2459094.835 & 0.005 & 0.055 & 1.092 & 60.9976329 & 28.6198327 & 13.162 & no \\
2018egj  & 18abeewyu & SN II & 2458303.653 & 0.308 & -0.943 & 0.157 & 250.955014 & 47.4085778 & 13.347 & no \\
2019wvz  & 19acytcsg & SN II & 2458831.951 & 0.005 & 0.102 & 1.986 & 155.119472 & 50.4679327 & 13.827 & no \\
2018fso  & 18abrlljc & SN II & 2458357.603 & 0.005 & -0.820 & 0.078 & 253.184108 & 70.0882348 & 14.107 & no \\
2019kes  & 19abegizf & SN II & 2458665.452 & 2.483 & -2.484 & 2.483 & 328.034016 & -23.364074 & 15.373 & no \\
2020aasd & 20actawpa & SN II & 2459175.306 & 0.437 & 0.604 & 0.714 & 140.812575 & 33.5669093 & 15.694 & no \\
2020sur  & 20abywoaa & SN II & 2459092.390 & 0.510 & -0.510 & 0.510 & 22.2482141 & -11.491487 & 16.537 & no \\
2020rid  & 20abpwdfd & SN II & 2459060.662 & 4.501 & 0.017 & 2.001 & 181.833961 & 57.6346927 & 16.729 & no \\
2018mbn  & 18abgxjie & SN II & 2458312.713 & 0.048 & 0.220 & 1.004 & 285.008531 & 51.9231897 & 17.981 & no \\
2019hln  & 19aaymhay & SN II & 2458642.772 & 0.011 & -0.805 & 0.010 & 287.957064 & 50.8476251 & 18.058 & no \\
2018cyh  & 18abcezmh & SN II & 2458284.848 & 2.120 & 0.035 & 0.964 & 269.451874 & 40.0763824 & 18.152 & no \\
2019fkl  & 19aavbjfp & SN II & 2458617.228 & 0.478 & -0.479 & 0.478 & 186.753673 & 62.1638376 & 18.569 & no \\
2020lcc  & 20abbeoaa & SN II & 2458991.762 & 0.004 & -5.973 & 0.030 & 231.265297 & 8.4907814 & 18.738 & no \\
2018bdv  & 18aapifti & SN II & 2458230.834 & 0.002 & 0.853 & 2.844 & 177.017646 & 30.3600716 & 18.883 & no \\
2019fmv  & 19aavbkly & SN II & 2458618.236 & 0.481 & -0.481 & 0.480 & 187.390843 & 35.7700404 & 20.564 & no \\
2020oco  & 20abjuxoy & SN II & 2459025.703 & 0.836 & 4.057 & 5.187 & 292.040677 & 52.8939923 & 22.092 & no \\
2018dzo  & 18abeteea & SN II & 2458303.683 & 0.050 & -0.843 & 0.027 & 230.63751 & 36.7986147 & 22.317 & no \\
2019pkh  & 19abuzinv & SN II & 2458726.450 & 0.520 & -0.520 & 0.520 & 34.8642701 & 34.0819977 & 22.550 & no \\
2020ttu  & 20acaiztt & SN II & 2459109.845 & 0.023 & -1.905 & 0.005 & 41.8004692 & 41.4099111 & 25.270 & no \\
2019jhe  & 19aaxqwjx & SN II & 2458639.712 & 0.002 & -0.001 & 0.045 & 236.891446 & 33.5519041 & 26.142 & no \\
2020umb  & 20acedspv & SN II & 2459116.625 & 0.090 & -0.875 & 0.125 & 336.870395 & 12.4784784 & 27.250 & no \\
2019aaqx & 19abmxtrm & SN II & 2458696.792 & 0.015 & -0.015 & 0.014 & 243.546215 & 59.0099127 & 32.209 & no \\
2018bjh  & 18aahrzrb & SN II & 2458217.581 & 0.044 & -0.841 & 0.159 & 181.397225 & 34.3888042 & 34.236 & no \\
2020drl  & 20aarbvub & SN II & 2458905.178 & 0.478 & -0.478 & 0.478 & 112.047976 & 72.5781035 & 34.456 & no \\
2020yyo  & 20acpevli & SN II & 2459157.715 & 0.002 & -0.682 & 0.197 & 167.480755 & 79.0043096 & 36.939 & no \\
2018lti  & 18abddjpt & SN II & 2458294.706 & 0.036 & -0.876 & 0.085 & 278.704811 & 38.2987135 & 40.294 & no \\
2018efj  & 18abimhfu & SN II & 2458320.653 & 0.002 & -0.903 & 0.017 & 240.142272 & 31.6429506 & 42.014 & no \\
2018mdz  & 18abcqhgr & SN II & 2458290.743 & 0.389 & 0.119 & 1.041 & 254.818204 & 60.4317906 & 46.105 & no \\
2020ks   & 20aaczkyw & SN II & 2458850.736 & 0.045 & -0.761 & 0.236 & 84.6766286 & 81.2087595 & 54.264 & no \\
2018cfj  & 18aavpady & SN II & 2458256.114 & 0.107 & -0.138 & 0.816 & 273.003116 & 44.3601877 & 55.886 & no \\
2020sfy  & 20abwftit & SN II & 2459085.460 & 0.480 & -0.480 & 0.480 & 33.103797 & -12.375219 & 58.527 & no \\
2018mdx  & 18aaxwrjt & SN II & 2458273.777 & 0.007 & -0.837 & 0.013 & 260.363393 & 25.6504719 & 70.223 & no \\
2020pvg  & 20abojbrd & SN II & 2459049.396 & 0.496 & -0.497 & 0.496 & 342.184072 & -19.829618 & 94.456 & no \\
\hline
\end{tabular}
\end{table*}


\newpage

\begin{table*}[h!]
\caption{Real infants classified as SN IIn }
\hspace{-3cm}
\begin{tabular}{lllllllllll}
\hline
IAU & ZTF & Type & Explosion & Error & Last & First & RA & DEC & First & Flasher \\
name & name &  & JD date &  & Non  & detection & (median) & (median) & spectrum &  \\
& (ZTF) &  &  &  & detection &  &  &  &  &  \\
& &  & [d] & [d] & [d] & [d] & [degrees] & [degrees] & [d] &  \\
\hline
\hline
2019njv & ZTF19abpidqn & SN IIn & 2458707.708 & 0.067 & -0.816 & 0.005 & 304.988297 & 15.3774528 & 1.150 & no \\
2020dcs & ZTF20aaocqkr & SN IIn & 2458895.153 & 0.032 & -0.168 & 0.778 & 183.356159 & 37.6993902 & 2.484 & yes \\
2020rfs & ZTF20abrmbdl & SN IIn & 2459072.375 & 0.445 & -0.445 & 0.445 & 282.303159 & 74.3340239 & 3.125 & no \\
2020xkx & ZTF20acklcyp & SN IIn & 2459137.743 & 1.008 & -1.008 & 1.007 & 350.11738 & 22.9869004 & 7.870 & no \\
2019smj & ZTF19aceqlxc & SN IIn & 2458767.998 & 0.970 & -0.970 & 0.969 & 117.419661 & 5.0742059 & 11.899 & no \\
2019dvw & ZTF19aapafqd & SN IIn & 2458571.416 & 1.512 & -1.513 & 1.512 & 239.944168 & 37.033706 & 15.539 & no \\
2019pgu & ZTF19abulzhy & SN IIn & 2458722.562 & 0.077 & 0.235 & 1.234 & 244.678475 & 67.9000902 & 18.120 & no \\
2020cnv & ZTF20aahapgw & SN IIn & 2458861.167 & 0.503 & -0.504 & 0.503 & 62.6199901 & 34.112782 & 30.834 & no \\
2018gfx & ZTF18abtswjk & SN IIn & 2458366.563 & 0.212 & -0.567 & 0.384 & 38.2980122 & -1.3056566 & 31.380 & no \\
2018dfa & ZTF18abcfdzu & SN IIn & 2458286.576 & 0.214 & -0.770 & 0.152 & 230.217161 & 54.2155543 & 32.424 & no \\
2019pdm & ZTF19abmouqp & SN IIn & 2458693.887 & 0.028 & 1.071 & 2.020 & 353.667326 & 16.4185618 & 39.113 & no \\
\hline
\end{tabular}
\label{tab:allSNIIn}
\end{table*}

\newpage

\begin{table*}[h!]
\caption{Real infants classified as SN IIb}
\hspace{-3cm}
\begin{tabular}{lllllllllll}
\hline
IAU & ZTF & Type & Explosion & Error & Last & First & RA & DEC & First & Flasher \\
name & name &  & JD date &  & Non  & detection & (median) & (median) & spectrum &  \\
& (ZTF) &  &  &  & detection &  &  &  &  &  \\
& &  & [d] & [d] & [d] & [d] & [degrees] & [degrees] & [d] &  \\
\hline
\hline
2020sbw & ZTF20abwzqzo & SN IIb & 2459087.465 & 0.482 & -0.482 & 0.482 & 41.5138221 & 3.329908 & 0.578 & no \\
2018dfi & ZTF18abffyqp & SN IIb & 2458307.254 & 0.432 & -0.432 & 0.432 & 252.708677 & 45.3978958 & 0.596 & yes \\
2018fzn & ZTF18abojpnr & SN IIb & 2458350.935 & 0.003 & -0.001 & 0.724 & 297.487196 & 59.5927746 & 0.962 & no \\
2019dwf & ZTF19aarfkch & SN IIb & 2458592.665 & 0.013 & -0.745 & 0.015 & 221.131644 & 70.4559895 & 1.183 & no \\
2019ehk & ZTF19aatesgp & SN IIb & 2458602.285 & 0.500 & -0.500 & 0.500 & 185.733956 & 15.826127 & 1.474 & yes \\
2019rwd & ZTF19acctwpz & SN IIb & 2458761.165 & 0.485 & -0.485 & 0.485 & 2.691207 & 21.1390942 & 1.568 & no \\
2020urc & ZTF20acgiglu & SN IIb & 2459123.366 & 0.474 & -0.475 & 0.474 & 34.5461651 & 37.0971887 & 2.341 & no \\
2018mdy & ZTF18aaymsbe & SN IIb & 2458276.640 & 0.005 & -1.756 & 0.166 & 243.77589 & 62.3191198 & 9.360 & no \\
2018jak & ZTF18acqxyiq & SN IIb & 2458442.985 & 0.960 & -0.960 & 0.959 & 149.825817 & 34.8954985 & 14.033 & no \\
2018efd & ZTF18abgrbjb & SN IIb & 2458312.867 & 0.005 & 0.066 & 0.850 & 274.998606 & 51.7964817 & 14.964 & no \\
\hline
\end{tabular}
\label{tab:allSNIIb}
\end{table*}

\newpage



\begin{figure}
    \hspace{-1cm}
    \includegraphics[trim = 80 20 80 0, clip, scale = 0.32]{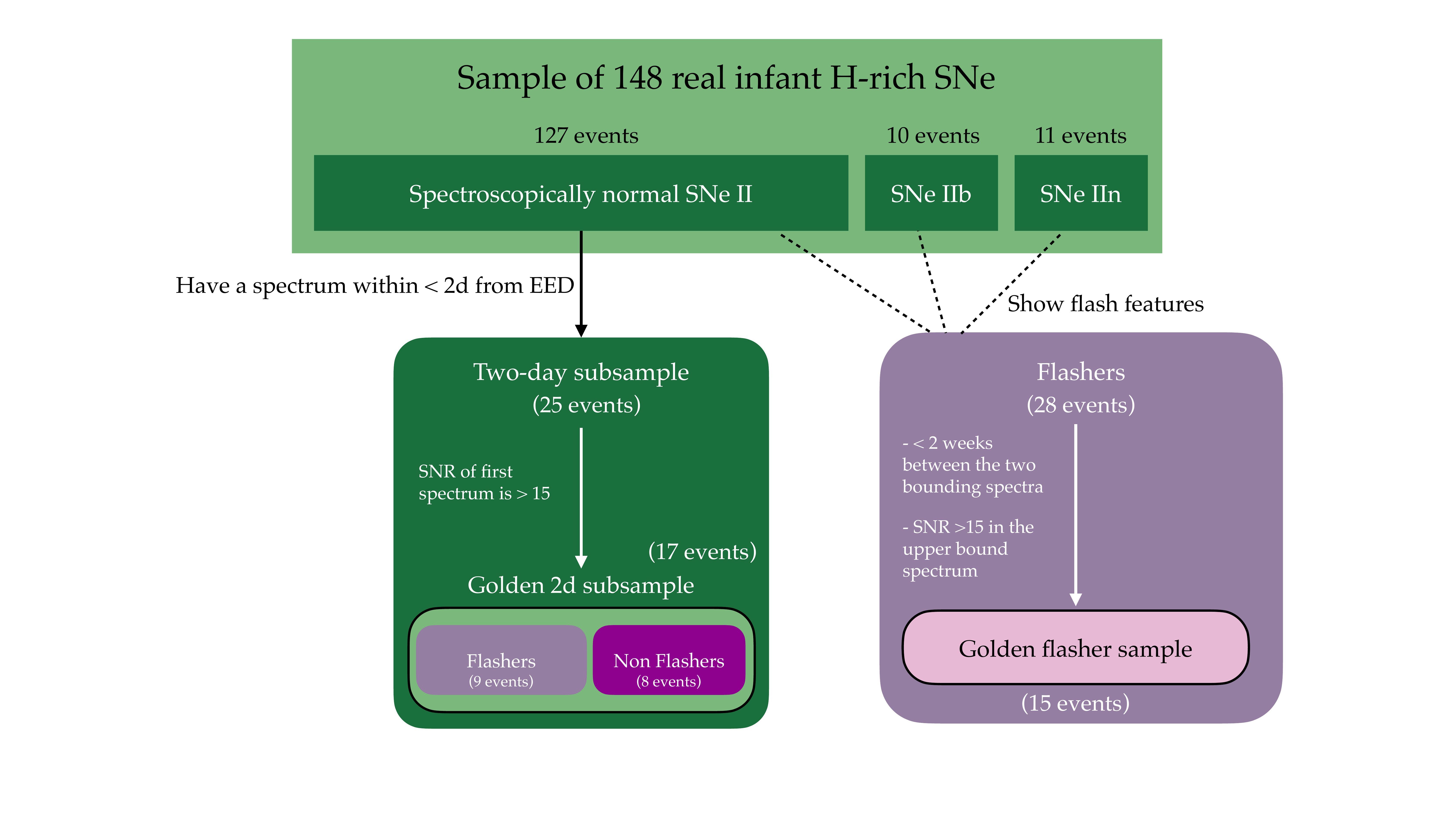}
    \caption{Schematic explanation of the subsamples used in this study. We use the golden 2-day subsample to compare the photometric parameters of flashers and non-flashers. We use the golden flasher sample to derive the durations of flash ionisation features }
    \label{fig:subsamps}
\end{figure}

\end{document}